\documentclass[%
 reprint,
 superscriptaddress,
 amsmath,amssymb,
 aps,
 prl,
 longbibliography
]{revtex4-2}

\usepackage{enumitem}
\usepackage[
 colorlinks=true,
 citecolor=blue,
 linkcolor=blue
]{hyperref}
\usepackage[pdftex]{graphicx} 

\makeatletter
\newcommand{\DisableTOC}{%
  \let\saved@addcontentsline\addcontentsline
  \renewcommand{\addcontentsline}[3]{}%
}
\newcommand{\EnableTOC}{%
  \let\addcontentsline\saved@addcontentsline
}
\makeatother

\usepackage{braket}
\usepackage{times,multirow,amsfonts,bm,xspace,pifont,mathtools}
\usepackage{soul}
\usepackage{ulem}
\usepackage{mathrsfs}
\usepackage{array}
\usepackage[T1]{fontenc}
\usepackage[utf8]{inputenc}

\begin{document}
\newcommand{\rr}{{\bm r}}
\newcommand{\q}{{\bm q}}
\renewcommand{\k}{{\bm k}}
\newcommand*\TKB[1]{\textcolor{blue}{#1}}
\newcommand{\TKBS}[1]{\textcolor{blue}{\sout{#1}}}
\newcommand*\YYWC[1]{\textcolor{green}{#1}}
\newcommand{\TK}[1]{{\color{red}{#1}}}
\newcommand*\TKS[1]{\textcolor{red}{\sout{#1}}}

\newcommand{\TKM}[1]{{\color{magenta}{#1}}}
\newcommand*\TKMS[1]{\textcolor{magenta}{\sout{#1}}}

\newcommand*\BdG{{\rm BdG}}
\newcommand{\mA}{\mathcal{A}}

\title{
Quantum-geometry-driven exact ferromagnetic ground state in a nearly flat band
}

\author{Taisei Kitamura}
\email[]{taisei-kitamura@g.ecc.u-tokyo.ac.jp}
\affiliation{Department of Physics, Graduate School of Science, The University of Tokyo, Tokyo 113-0033, Japan}
\affiliation{RIKEN Center for Emergent Matter Science (CEMS), Wako 351-0198, Japan}

\author{Hiroki Nakai}
\affiliation{Graduate School of Arts and Sciences, The University of Tokyo, Tokyo 153-8902, Japan}
\affiliation{Department of Physics, University of Toronto, 60 St. George St., Toronto, Ontario, M5S 1A7, Canada}

\author{Hosho Katsura}
\affiliation{Department of Physics, Graduate School of Science, The University of Tokyo, Tokyo 113-0033, Japan}
\affiliation{Institute for Physics of Intelligence, The University of Tokyo, Tokyo 113-0033, Japan}
\affiliation{Trans-Scale Quantum Science Institute, The University of Tokyo, Tokyo 113-0033, Japan}

\author{Ryotaro Arita}
\affiliation{Department of Physics, Graduate School of Science, The University of Tokyo, Tokyo 113-0033, Japan}
\affiliation{RIKEN Center for Emergent Matter Science (CEMS), Wako 351-0198, Japan}

\date{\today}

\begin{abstract}

We construct a Hubbard model with a nearly flat band whose quantum geometry can be tuned independently of the energy dispersion and the Coulomb interaction. We show that, when the nearly flat band is half-filled, the exact ground state of the model exhibits ferromagnetism and that this ferromagnetism is stabilized by the quantum metric through the spin stiffness. Furthermore, we demonstrate that tuning the quantum geometry alone drives a magnetic phase transition. Our nonperturbative results without resorting to mean-field approximations reveal the quantum‑geometric origin of ferromagnetism and the underlying many-body physics in dispersive-band systems.
\end{abstract}

\maketitle

\DisableTOC 

\textit{Introduction---}Strongly correlated electron systems serve as a platform for quantum phases of matter, such as magnetism and superconductivity. 
The Hubbard model~\cite{Kanamori1963,Gutzwiller1963,Hubbard1963} is a minimal model that describes quantum many-body physics underlying these phenomena.
The interplay between the Bloch electrons and the Coulomb interaction leads to various types of quantum phases.
For example, the half-filled Hubbard model can exhibit a Mott insulating state due to the competition between kinetic-energy gain and Coulomb repulsion~\cite{Georges1996,Imada1998,Kotliar2004-zw,Lee2006-qx}.
On the square lattice, carrier doping can promote unconventional $d$-wave superconductivity, often discussed in terms of spin-fluctuation mechanisms~\cite{Moriya2000-dn,Yanase2003,Monthoux2007-ew,Scalapino2012}.

In a canonical single-band case, the Bloch wave function is given by simple plane waves with a trivial momentum-space property.
Accordingly, an energetic picture based solely on the kinetic energy and the Coulomb repulsion is applicable.
Moreover, the emergent phenomena of quantum phases, such as magnon excitation~\cite{Izuyama1963} and the Meissner effect~\cite{Scalapino1992,Scalapino1993,Jujo2001}, are characterized by the energy dispersion and its effective mass.
In contrast, the Bloch wave function of multiband systems manifests nontrivial geometric properties in momentum space, that is, quantum geometry~\cite{Provost1980,Resta2011}. Recent years have witnessed the essential role of quantum geometry in quantum phases of strongly correlated electron systems~\cite{Rossi2021,Torma2022,Torma2023,Yu2025,Kitamura2025perspective}.

Flat-band systems offer a paradigmatic platform for strongly correlated electrons, where quantum geometry plays a key role, and moiré materials provide a primary route to realizing such bands~\cite{Andrei2021,Nuckolls2024}.
A well-known instance is flat-band superconductivity, where quantum geometry gives rise to the superfluid weight and thereby induces the Meissner effect~\cite{Peotta2015,Liang2017,Huhtinen2022,Peotta2023}.
Indeed, the quantum-geometric origin of the superfluid weight in twisted multilayer graphene~\cite{Cao2018,Hao2021,Park2021} has been intensively discussed theoretically~\cite{Hu2019,Xie2020,Julku2020,Hirobe2025} and experimentally~\cite{Tian2023,Tanaka2025,Banerjee2025}.
Moreover, quantum geometry is essential for flat-band magnetism~\cite{Wu2020,Kang2024, Han2024,Zhang2025,Kitamura2025,Pichler2025}, as it ensures the uniqueness of the ground state~\cite{Kitamura2025} in flat-band ferromagnetism~\cite{Lieb1989,Mielke1991,Mielke1991further,Mielke1992,Tasaki1992,Mielke1993,Tasaki1995,Mielke1999,Katsura2010,Pons2020,Tasaki2020,Gulacsi2007-xr,Trencsenyi2011-zq,Gulacsi2014-bx,Gulacsi2014-wg,Arita2002-xg,Suwa2003-dg}.

Beyond the flat-band limit, quantum geometry can drive various superconducting~\cite{Ahn2021,Chen2021,Kitamura2022,Kitamura2023,Kitamura2024,Yu2024,Daido2024,Zhang2024,Simon2025} and magnetic phenomena~\cite{Kitamura2024,Heinsdorf2025,Kitamura2025,Kudo2025,Hu2025,Oh2025,Espinosachampo2025,Wang2025,Nogaki2025} in dispersive-band systems.
Notably, it is revealed that the competition between quantum geometry and band dispersion strongly impacts magnetic phase diagrams~\cite{Kitamura2024,Kitamura2025,Hu2025,Oh2025}.
Considering that an experimentally relevant flat band possesses a finite bandwidth and, thereby, is a nearly flat, like twisted bilayer graphene~\cite {Suarez2010,Bistritzer2011,Cao2018}, this competition may be pertinent to realistic flat band materials~\cite{Hu2025}. 
Therefore, magnetism in dispersive-band systems, including nearly flat-band systems, is a growing body of research on quantum geometry~\cite{Kitamura2025perspective}.

However, investigations of dispersive-band systems, especially regarding magnetism, are mostly based on the mean-field approximation, leaving the quantum-geometry-driven phases and underlying quantum many-body physics uncertain.
This is in sharp contrast to the system with a perfect flat-band, in which analysis beyond the mean-field framework can be carried out~\cite{Herzog2022,Herzog2022prl,Hofmann2023,Han2024,Zhang2025,Kitamura2025,Han2025}.
Considering that mean-field theories often fail to describe the physics of itinerant magnetism~\cite{Kanamori1963,Moriya1973,Moriya1985}, elucidating quantum-geometry-driven magnetism in dispersive-band systems beyond the mean-field framework is an important open task.

In this Letter, we present an example of quantum-geometry-driven magnetism in dispersive-band systems, with the aid of rigorous theories~\cite{Tasaki1995,Tanaka2001,Tanaka2003,Ueda2004,Lu2009,Tanaka2018,Tamura2019} and without resorting to mean-field approximations.
We introduce a class of Hubbard models on one-dimensional (1D) lattices with tunable quantum geometry that exhibit the exact ferromagnetic ground state in the lowest nearly flat band.
To reveal the quantum geometric effect, we formulate the spin-excitation energy and the spin stiffness in terms of the energy dispersion and the quantum geometry of Bloch electrons.
These formulas show that the quantum metric stabilizes ferromagnetism and that tuning the quantum geometry alone leads to a magnetic phase transition by keeping the energy dispersion and the Coulomb interaction fixed.
Our results cannot be captured within the mean-field framework, thereby highlighting the essential role of quantum geometry in quantum many-body physics of dispersive band systems.

\begin{figure}[tbp]
  \includegraphics[width=1.0\linewidth]{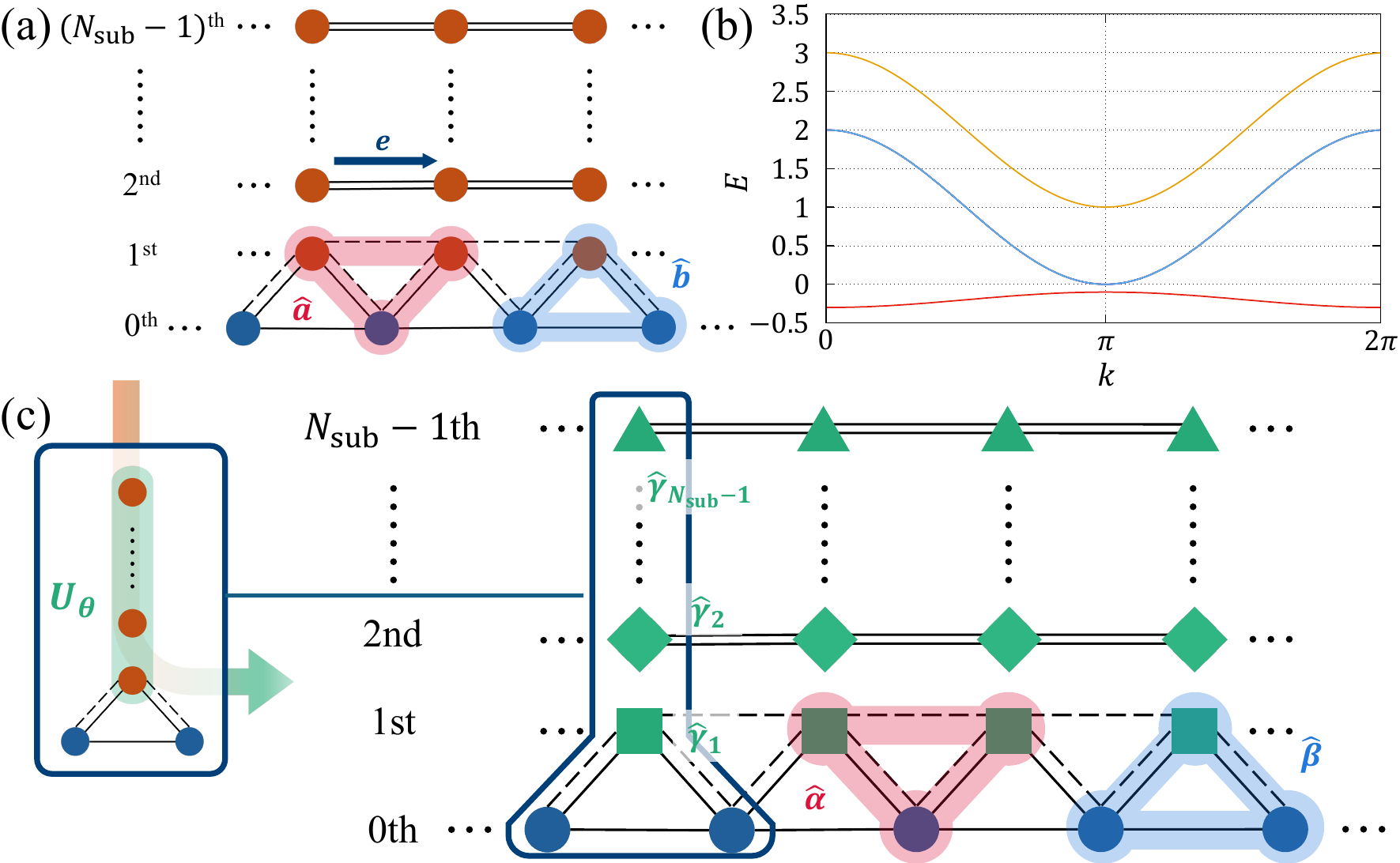}
  \centering
  \caption{(a) Tasaki's delta chain and 1D periodic chains with nearest neighbor hoppings. The lattice vector is represented by the navy-blue arrow. (b) Energy dispersion of $\hat{H}_{0}$ for $(s,t,u,v,\lambda) = (1/20,1/2,1/2,1,\sqrt{2})$. The middle bands are $(N_{\rm sub}-2)$-fold degenerate. (c) The unitary transformation from $\hat{H}_{0:\rm T} + \hat{H}_{0:\rm NN}$ to $\hat{H}_{0}$ by $\mathcal{U}_\theta$ is schematically illustrated.}
  \label{fig:bcc_schematic}
\end{figure}

\textit{Hubbard model on 1D lattices with tunable quantum geometry---}We consider a set of 1D periodic chains, each containing $N$ sites and aligned along the lattice vector $\bm e=(1,0,0)$. The total number of chains is $N_{\rm sub}$, and the sites on the $i$th chain are located at positions $\bm r = \bm x + \bm \mu_i$ for $i=0,1,\dots,N_{\rm sub}-1$. Here, we define $\bm x = x\,\bm e$ with $x\in\{0,1,\dots,N-1\}$. Each chain is shifted from the reference chain by a displacement vector $\bm \mu_i$.
We set $\bm \mu_0 = 0$ as the $0$th chain lies on $\bm x$ and is the reference chain.
For $i\neq0$, the displacement along the lattice vector is fixed at half the lattice spacing, i.e., $[\bm \mu_i]_1 = 1/2$. This configuration is schematically illustrated in Fig.~\ref{fig:bcc_schematic} (a), where the filled blue and orange circles denote the reference and the other chains, respectively.

Given this setup, we construct a Hubbard model with tunable quantum geometry that is rigorously shown to exhibit ferromagnetism.
Tasaki proved that the ferromagnetic ground state of the flat-band Hubbard model on the delta chain~\cite{Tasaki1992} persists even when the flat band becomes dispersive due to appropriate extra hoppings, provided that the Coulomb interaction is sufficiently large~\cite{Tasaki1995}. This is the starting point of our model construction.
We introduce an auxiliary Hamiltonian composed of Tasaki’s delta chain together with $(N_{\rm sub}-2)$ additional 1D chains with nearest neighbor (NN) hoppings as follows. 

Tasaki's delta chain is formed by the $0$th and the $1$st chains and is written in terms of the $a$ operators
$
\hat{a}_{\bm x, \sigma}= \lambda \hat{c}_{\bm x, \sigma}-(\hat{c}_{\bm x+\bm \mu_1, \sigma}+\hat{c}_{\bm x+\bm \mu_1-\bm e, \sigma})
$
and
the $b$ operators
$
\hat{b}_{\bm x, \sigma}= \hat{c}_{\bm x, \sigma}+\lambda \hat{c}_{\bm x+\bm \mu_1, \sigma}+\hat{c}_{\bm x+\bm e, \sigma}
$, which are shown by the red downward triangle and blue upward triangle in Fig.~\ref{fig:bcc_schematic} (a), 
respectively.
Here, $\hat{c}_{\bm r, \sigma}$ is the fermionic annihilation operator at site $\bm r$ with spin $\sigma = \uparrow,\downarrow$, and the positive $\lambda$ controls the weight distribution in the $a$ and $b$ operators.
In Ref.~\cite{Tasaki1992}, Tasaki showed that the hopping Hamiltonian $t\sum_{\bm x,\sigma}\hat{b}_{\bm x,\sigma}^\dagger\hat{b}_{\bm x,\sigma}$ with positive $t$ possesses an isolated lowest flat band (see Fig.~\ref{fig:bcc_schematic}(a) in which hoppings shown by solid lines form a delta chain).
Since $\hat{b}_{\bm x,\sigma}$ anticommutes with $\hat{a}_{\bm x,\sigma}^\dagger$, $\hat{a}_{\bm x, \sigma}^\dagger\ket{0}$ with vacuum state $\ket{0}$ are flat-band eigenstates.
To make this flat band dispersive, we add extra hopping terms following Ref.~\cite{Tasaki1995}; the resulting Hamiltonian takes the form $\hat{H}_{0:{\rm T}} = \sum_{\bm x,\sigma}[-s\hat{a}_{\bm x,\sigma}^\dagger\hat{a}_{\bm x,\sigma}+t\hat{b}_{\bm x,\sigma}^\dagger\hat{b}_{\bm x,\sigma}]$, where we take $s>0$. 
The extra hopping terms proportional to $s$ form the inverted delta chain, as shown by the dashed line in Fig.~\ref{fig:bcc_schematic} (a).

The $a$ and $b$ operators satisfy the following anticommutation relations: $\{\hat{a}_{\bm x,\sigma}^\dagger,\hat{a}_{\bm x,\sigma}\} = \{\hat{b}_{\bm x,\sigma}^\dagger, \hat{b}_{\bm x,\sigma}\} = \lambda^2+2$, $\{\hat{a}_{\bm x,\sigma}^\dagger,\hat{a}_{\bm x+\bm e,\sigma}\} = \{\hat{b}_{\bm x,\sigma}^\dagger,\hat{b}_{\bm x+\bm e,\sigma}\} = 1$, and all other anticommutators vanish.
Consequently, within $\hat{H}_{0:{\rm T}}$, the single-particle states generated by the $a$ and $b$ operators obey the standard 1D Schr\"odinger equation. For the $a$ operators, the NN hopping and the onsite potential are given by $-s$ and $-s(\lambda^2+2)$, yielding the energy dispersion $\epsilon_{\alpha}(k)=-s[\lambda^2+2+2\cos k]$ with momentum $k$. Similarly, the energy dispersion corresponding to the $b$ operators is found to be $\epsilon_{\beta}(k)=t[\lambda^2+2+2\cos k]$.

The remaining 1D chains with NN hoppings are described by
$\hat{H}_{0:\rm{NN}} = \sum_{i(\neq0,1)}\sum_{\bm x,\sigma}[v\hat{c}_{\bm x+\bm \mu_i,\sigma}^\dagger\hat{c}_{\bm x+\bm \mu_i,\sigma}+ u(\hat{c}_{\bm x+\bm \mu_i+\bm e,\sigma}^\dagger\hat{c}_{\bm x+\bm \mu_i,\sigma}+{\rm H.c.})]$ with $v>0$ and $u>0$.
This is schematically illustrated by double solid lines in Fig.~\ref{fig:bcc_schematic} (a).
The single-particle eigenvalues of $\hat{H}_{0:\rm{NN}}$, each $(N_{\rm sub}-2)$-fold degenerate, are given by $v+2u\cos k$~\cite{degenerate}.
By adding the onsite Coulomb interaction $\hat{H}_{\rm int} = U\sum_{\bm r}\hat{n}_{\bm r,\uparrow}\hat{n}_{\bm r,\downarrow},$ with $U>0$, the auxiliary Hubbard Hamiltonian reads $\hat{H}_{0:{\rm T}}+\hat{H}_{0:{\rm NN}} + \hat{H}_{\rm int}$.
By Tasaki's theorem~\cite{Tasaki1995}, we can prove that the half-filled lowest band $\epsilon_{\alpha}(k)$ exhibits ferromagnetism for sufficiently large $t/s,(v-2u)/s,$ and $U/s$.

In our auxiliary Hubbard Hamiltonian, the delta chain and the remaining 1D chains are completely decoupled.
We couple them without altering the interaction term by mixing the operators of the chains from the $1$st to the $(N_{\rm sub}-1)$th ones via a unitary transformation:
$
\hat{\bm \gamma}_{\bm x,\sigma} = [1\oplus \mathcal{U}_{\theta}] \hat{\bm c}_{\bm x,\sigma}
$, where $\hat{\bm \gamma}_{\bm x,\sigma} = (\hat{\gamma}_{0,\bm x,\sigma},\cdots,\hat{\gamma}_{N_{\rm sub}-1,\bm x,\sigma})^{\mathsf T}$ and $\hat{\bm c}_{\bm x,\sigma} = (\hat{c}_{\bm x+\bm \mu_0,\sigma},\cdots,\hat{c}_{\bm x+\bm \mu_{N_{\rm sub}-1},\sigma})^{\mathsf T}$.
Here, $\mathcal{U}_\theta$ is an arbitrary unitary matrix that depends continuously on $\theta$ (see Eq.~\eqref{eq:Utheta} for an explicit example), and its unitarity ensures that the $\gamma$ operators satisfy the canonical anticommutation relations, i.e., $\{\hat{\gamma}_{i,\bm x,\sigma}^\dagger,\hat{\gamma}_{j,\bm x',\sigma'}\} = \delta_{i,j}\delta_{\bm x,\bm x'}\delta_{\sigma,\sigma'}$.
We then introduce the noninteracting Hamiltonian constructed by the delta chain and $(N_{\rm sub}-2)$ 1D chains with NN hoppings in terms of $\hat{\bm \gamma}_{\bm x,\sigma}$ as
\begin{align}
&\hat{H}_{0}(\theta) = \sum_{\bm x,\sigma}\Big[-s\hat{\alpha}_{\bm x,\sigma}^\dagger\hat{\alpha}_{\bm x,\sigma}
+t\hat{\beta}_{\bm x,\sigma}^\dagger\hat{\beta}_{\bm x,\sigma}
\notag\\
&+\sum_{i(\neq 0,1)}\{v\hat{\gamma}_{i,\bm x,\sigma}^\dagger\hat{\gamma}_{i,\bm x,\sigma}
+u(\hat{\gamma}_{i,\bm x+\bm e,\sigma}^\dagger\hat{\gamma}_{i,\bm x,\sigma}+h.c.)\}\Big],    
\end{align}
with 
$
\hat{\alpha}_{\bm x, \sigma}= \lambda \hat{\gamma}_{0, \bm x, \sigma}-(\hat{\gamma}_{1,\bm x, \sigma}+\hat{\gamma}_{1,\bm x-\bm e, \sigma})
$
and
$
\hat{\beta}_{\bm x, \sigma}= \hat{\gamma}_{0, \bm x, \sigma}+\lambda \hat{\gamma}_{1,\bm x, \sigma}+\hat{\gamma}_{0, \bm x+\bm e, \sigma}
$.
This noninteracting Hamiltonian depends on $\theta$ through the $\gamma$ operators.
See Fig.~\ref{fig:bcc_schematic} (c) for a schematic illustration of the unitary transformation where $\alpha$, $\beta$, and $\gamma$ operators are represented by the red downward triangle, the blue upward triangle, and the green shapes, respectively.
By adding the Coulomb interaction term, our Hubbard Hamiltonian is defined by $\hat{H}(\theta) = \hat{H}_0(\theta)+\hat{H}_{\rm int}$. 
In the following, we suppress $\theta$ for notational simplicity.

$\hat{H}_0$ is unitarily equivalent to $\hat{H}_{0:{\rm T}} + \hat{H}_{0:{\rm NN}}$ and therefore 
has the same energy dispersion, which is shown in Fig.~\ref{fig:bcc_schematic}(b) for $(s,t,u,v,\lambda) = (1/20,1/2,1/2,1,\sqrt{2})$.
However, due to the mixing by $\mathcal{U}_\theta$, the hopping processes in the sublattice basis become complicated and depend on $\theta$, indicating that the Bloch wave function also depends on $\theta$. Therefore, without altering energy dispersion and the strength of the Coulomb interaction, we can tune the quantum geometry via $\theta$. 
This is crucial for many-body physics mediated by the Coulomb interaction since changing $\mathcal{U}_\theta$ modifies the matrix elements of $\hat{H}_{\rm int}$ in the Bloch-electron basis.
In other words, the effective Coulomb interaction between Bloch electrons is modulated by $\theta$ through quantum geometry.

\textit{Saturated ferromagnetism in a nearly flat band---}Here, we prove the saturated ferromagnetism of the half-filled lowest band $\epsilon_\alpha(k)$.
Because of the $SU(2)$ symmetry, eigenstates are labeled by the eigenvalue of $(\hat{\bm S}_{\rm tot})^2$, i.e., $S_{\rm tot}(S_{\rm tot}+1)$ where we define $\hat{\bm S}_{\rm tot}= \sum_{\bm r}\hat{\bm c}_{\bm r}^\dagger \bm \sigma \hat{\bm c}_{\bm r}/2$ with $\hat{\bm c}_{\bm r} = (\hat{c}_{\bm r,\uparrow}, \hat{c}_{\bm r,\downarrow})$ and $\bm \sigma = (\sigma_1,\sigma_2,\sigma_3)$, the vector of Pauli matrices.
If the total spin of any ground state takes the maximum value, we say that the model exhibits saturated ferromagnetism~\cite{Tasaki2020}.

Following the strategy developed by Tasaki and others~\cite{Tasaki1995,Tanaka2001,Tanaka2003,Ueda2004,Lu2009,Tanaka2018,Tamura2019}, we fix the total electron number to $N$ and set $v-2u>0$ to ensure that $\epsilon_\alpha(k)$ is the lowest band. Within the Stoner theory, the 1D half-filled cosine dispersion of $\epsilon_\alpha(k)$ exhibits perfect nesting at the $\pi$ point, resulting in a strong antiferromagnetic instability. For completeness, we show the Stoner factor indicating antiferromagnetic order in Appendix A.
However, in the following, saturated ferromagnetism is established in a purely nonperturbative and rigorous manner.

Our starting point is the decomposition of the Hamiltonian into a flat-band Hamiltonian and local terms:
$\hat{H} = \xi\hat{H}_{\rm flat}+\sum_{\bm x} \hat{h}_{\bm x}$, where $\xi$ is a positive infinitesimal.
Here, the flat-band Hamiltonian is defined by $\hat{H}_{\rm flat} \equiv \hat{H}\vert_{s=0}$ whose lowest band is perfectly flat. 
The local Hamiltonian $\hat{h}_{\bm x}$ is chosen to be translation invariant and to have a saturated ferromagnetic state of the lowest band $\ket{\hat{\psi}_{\rm all\uparrow}}= \prod_{\bm x}\hat{\alpha}_{\bm x,\uparrow}^\dagger\ket{0}$ with $S_{\rm tot} = N/2$ as an eigenstate.
Clearly, $\ket{\hat{\psi}_{\rm all\uparrow}}$ is a ground state of $\hat{H}_{\rm flat}$. From Mielke's theorem for the Hubbard model with a flat lowest band~\cite{Mielke1999, Tasaki2020}, one can readily verify that $\ket{\hat{\psi}_{\rm all\uparrow}}$ is, up to the trivial spin degeneracy due to the $SU(2)$ symmetry, a unique ground state of $\hat{H}_{\rm flat}$. Thus, the flat-band Hamiltonian $\hat{H}_{\rm flat}$ exhibits saturated ferromagnetism.

The ground state of $\hat{H}_{\rm flat}$ is also an eigenstate of $\hat{H}$, i.e., $\hat{H}\ket{\hat{\psi}_{{\rm all}\uparrow}}  = N[-s(\lambda^2+2)]\ket{\hat{\psi}_{{\rm all}\uparrow}}$.  
By the comparison with $\hat{H}\ket{\hat{\psi}_{\rm all\uparrow}}=\sum_{\bm x}\hat h_{\bm x}\ket{\hat{\psi}_{\rm all\uparrow}}= N\hat h_{\bm x}\ket{\hat{\psi}_{{\rm all}\uparrow}}$ owing to $\hat{H}_{\rm flat}\ket{\psi_{\rm all\uparrow}} = 0$ and the translational invariance of $\hat h_{\bm x}$,
we obtain $\hat h_{\bm x}\ket{\hat{\psi}_{\rm all\uparrow}}=-s(\lambda^2+2)\ket{\hat{\psi}_{\rm all\uparrow}}$.
Let $\varepsilon_0$ be the minimum eigenvalue of $\hat{h}_{\bm x}$. If $\varepsilon_0 = -s (\lambda^2+2)$, then
$\ket{\hat{\psi}_{\rm all\uparrow}}$ is the simultaneous ground state of $\hat{H}_{\rm flat}$ and each local term $\hat{h}_{\bm x}$. In addition, since $\ket{\hat{\psi}_{\rm all\uparrow}}$ is the unique ground state of $\hat{H}_{\rm flat}$, $\ket{\hat{\psi}_{\rm all\uparrow}}$ is also the unique ground state of the total Hamiltonian $\hat{H}$.
In other words, when $\varepsilon_{0} = -s(\lambda^2+2)$, $\hat{H}$ exhibits saturated ferromagnetism.

By taking the limit of $t,v, U\rightarrow\infty$ with all other parameters fixed, we can analytically prove that $\varepsilon_0=-s(\lambda^2+2)$ for a certain choice of $\hat{h}_{\bm x}$ (see Appendix B). 
This, in turn, implies saturated ferromagnetism in the ground state for sufficiently large $t,v$ and $U$.
Here, to analyze finite parameter regions, we study the case of $N_{\rm sub} = 5$ and 
\begin{eqnarray}    
    \mathcal{U}_\theta = \left(
    \begin{array}{cccc}
       \cos^2\theta &
       \frac{1}{2}\sin2\theta  &
       \frac{1}{2}\sin2\theta & \sin^2\theta \\  
       -\sin\theta & \cos\theta & 0 & 0 \\  
       -\frac{1}{2}\sin2\theta & -\sin^2\theta & \cos^2\theta & \frac{1}{2}\sin2\theta \\  
       0 & 0 & -\sin\theta & \cos\theta
    \end{array}
    \right),\label{eq:Utheta}
\end{eqnarray}
and numerically diagonalize $\hat{h}_{\bm x}$ with its specific choice given in Appendix B.
This $\mathcal{U}_{\theta}$ is a minimal one-parameter family of special orthogonal matrices and continuously connects the decoupled-chain limit at $\theta = 0$ to the fully mixed one at $\theta = \pi/4$, where $\hat{\gamma}_{1}$ becomes an equal superposition of modes on the four chains~\cite{unitary}.
Considering that $\hat{\gamma}_{1}$ enters the localized state of the nearly flat band, $\theta$ directly controls the spread of the nearly-flat-band Wannier function.
In the following calculation, we set $\lambda = \sqrt{2}$, $s = 1/20$, $t = 30$, $u  = 5$, and $v= 40$ and take the limit $\xi\rightarrow+0$; the lowest band is isolated by a large band gap and is nearly flat compared with other bands.
In this setup, we study how large $U$ is required to achieve saturated ferromagnetism with a given quantum geometry, i.e., given $\theta$ while keeping the energy dispersion fixed.

\begin{figure}[tbp]
  \includegraphics[width=1.0\linewidth]{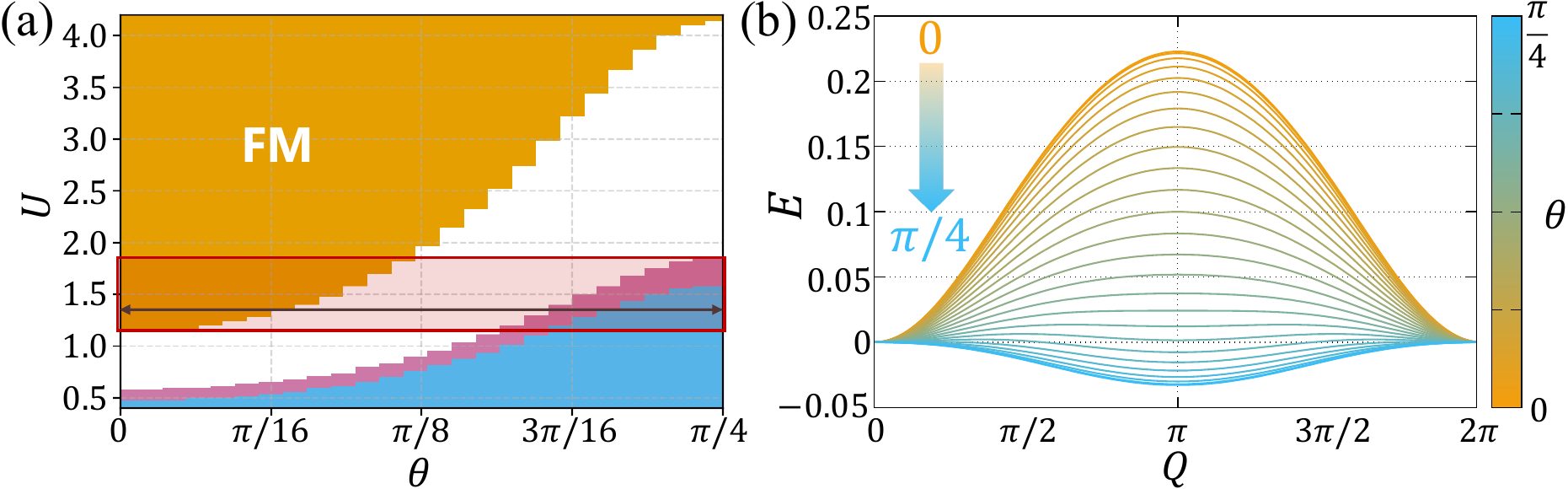}
  \centering
  \caption{(a) The orange region denotes where the ground state is saturated ferromagnetism, while blue and purple regions denote where the saturated ferromagnetic state is unstable. In the blue region, $\mathcal{D}_{\rm spin}<0$ is additionally satisfied. (b) Magnon dispersion for $U = 1.36$ and several values of $\theta$. The orange and blue colors correspond to $\theta = 0$ and $\theta = \pi /4$, respectively.}
  \label{fig:phase_diagram}
\end{figure}

Figure~\ref{fig:phase_diagram}(a) shows the result as a function of $U$ and $\theta$.
In the orange region, $\varepsilon_{0} = -s(\lambda^2+2)$ is satisfied, which indicates that the ground state exhibits saturated ferromagnetism.
We find that the strength of the onsite Coulomb interaction required for the saturated ferromagnetism strongly depends on $\theta$. This implies that a possible magnetic phase transition is driven by tuning the quantum geometry alone through the change of $\theta$.
However, $\varepsilon_{0}=-s(\lambda^2+2)$ is a sufficient but not necessary condition for saturated ferromagnetism~\cite{sufficient}.
In other words, the saturated ferromagnetism may still be realized even when $\varepsilon_{0}<-s(\lambda^2+2)$. Thus, we need an alternative method to show the instability of ferromagnetism.

\textit{Spin-excitation energy and spin stiffness---}The spin excitation energy of the many-body Hamiltonian is a criterion for the instability of ferromagnetism; the negative spin excitation energy indicates the instability of ferromagnetism~\cite{Kusakabe1994,Penc1996,Neupert2012,Tasaki1994,Tasaki1996,Wu2020,Kang2024}.
Because of the large energy gap between the lowest and the other bands, the low-energy spin-excitations are well approximated by single-spin-flip states from the saturated ferromagnetic state within the lowest band~\cite{Kusakabe1994,Penc1996,Neupert2012}.
Therefore, we consider the effective Hamiltonian projected onto the subspace spanned by these states.

To construct the effective Hamiltonian, it is useful to introduce the momentum space formalism.
We rewrite the noninteracting Hamiltonian as
$
\hat{H}_0 = \sum_{\sigma, k}\hat{\bm c}_\sigma^\dagger(k)H_0(k)\hat{\bm c}_\sigma(k)
$. 
Here, $\hat{\bm c}_\sigma(k)$ is the vector representation of the annihilation operator in momentum space~\cite{Supple}.
The matrix representation of the noninteracting Hamiltonian, $H_0(k)$, is diagonalized in the Bloch basis as $H_0 (k) \ket{u_n(k)} = \epsilon_n (k) \ket{u_n(k)}$, where $\ket{u_n(k)}$ is the periodic part of the Bloch wave function for band $n$. Note that $n = \alpha$ corresponds to the lowest band.
We denote by $\hat{d}^\dagger_{n,\sigma}(k)$ the operator that creates the orthonormal eigenstate of $\hat{H}_0$ corresponding to $\ket{u_n(k)}$. In terms of these operators, the normalized saturated ferromagnetic state in the lowest band can be rewritten as $\ket{\hat{\phi}_{\rm all\uparrow}} = \prod_{k}\hat{d}^\dagger_{\alpha,\uparrow}(k)\ket{0}$.

In this setup, the single-spin-flip state within the lowest band is defined by,
$
\ket{\hat{\phi}_{Q:k}}=\hat{d}_{\alpha,\downarrow}^\dagger(k+Q)\hat{d}_{\alpha,\uparrow}(k)\ket{\hat{\phi}_{\rm all\uparrow}}
$ with momentum $Q$.
Since $Q$ behaves as a good quantum number in the subspace spanned by $\ket{\hat{\phi}_{Q:k}}$, the effective Hamiltonian is labeled by $Q$ as $\mathcal{H}_{Q}$ and is defined via its matrix elements, $[\mathcal{H}_{Q}]_{k,k^\prime} = \bra{\hat{\phi}_{Q:k}}\hat{H}\ket{\hat{\phi}_{Q:k^\prime}}$.
We denote the orthonormal eigenvectors of $\mathcal{H}_{Q}$ by $\bm A_Q^{\chi} = (\cdots, A_{Q}^\chi(k),\cdots)^{\mathsf T}$, where $\chi =0,1,2\ldots,N-1$ labels the eigenvalues in ascending order, such that $\chi = 0$ indicates the lowest eigenvalue. 
Then, we define the spin-excitation energy by $E_{{\rm spin}}^\chi(Q) = \bra{\hat{\phi}_{Q}^\chi}\hat{H}\ket{\hat{\phi}_{Q}^\chi}-E_{\rm FM}$ with $
\ket{\hat{\phi}_{Q}^\chi}=\sum_{k}A_{Q}^\chi(k)\ket{\hat{\phi}_{Q:k}}
$ and $E_{\rm FM} = -Ns(\lambda^2+2)$ being the energy of the saturated ferromagnetic state. 
If $E_{{\rm spin}}^0(Q) < 0$ holds, then, by the variational principle, there exists at least one eigenvalue of $\hat{H}$ that is lower than $E_{\rm FM}$, ruling out the possibility that the saturated ferromagnetic state is the ground state. This, in turn, implies the instability of the saturated ferromagnetic state. Note, however, that this condition alone does not provide any information about the actual ground state.

In Fig.~\ref{fig:phase_diagram}(a), we plot the region where the lowest spin-excitation energy $E_{\rm spin}^0(Q)$ is negative in blue or purple; saturated ferromagnetism is unstable in these regions.
Moreover, in the blue region, the spin stiffness defined by $\mathcal{D}_{\rm spin} = \lim_{Q\rightarrow0}\partial_Q^{2}E_{\rm spin}^{0}(Q)$, is also negative, where the saturated ferromagnetism is not stable, even locally.
This is strong evidence that the ground state is not a ferromagnetic state.
For example, the parameter $U = 1.36$, marked by the horizontal arrow in Fig.~\ref{fig:phase_diagram}(a), indicates a magnetic phase transition driven by tuning quantum geometry via $\theta$ alone without any other changes.
This is also confirmed in Fig.~\ref{fig:phase_diagram}(b), where we show the magnon dispersion for several values of $\theta$. 
As $\theta$ increases from $0$ to $\pi/4$, $E_{\rm spin}^0(Q)$ decreases to $0$ and eventually becomes negative. 
Such a phase transition occurs over a wide range of $U$ ($1.2 \le U \le 1.8$), as highlighted in a red rectangle.

\textit{Quantum geometric origin of saturated ferromagnetism---}To interpret these results from the viewpoint of quantum geometry, we formulate $[\mathcal{H}_{Q}]_{k,k^\prime}$ and $\mathcal{D}_{\rm spin}$ in terms of the Bloch states.
First, by projecting $\hat{H}_{0}$ and $\hat{H}_{\rm int}$ onto the subspace spanned by $\ket{\hat{\phi}_{Q:k^\prime}}$ as $[\mathcal{H}_{Q}^{0}]_{k,k^\prime} = \bra{\hat{\phi}_{Q:k}}\hat{H_0}\ket{\hat{\phi}_{Q:k^\prime}}$ and $[\mathcal{H}_{Q}^{\rm int}]_{k,k^\prime} = \bra{\hat{\phi}_{Q:k}}\hat{H}_{\rm int}\ket{\hat{\phi}_{Q:k^\prime}}$, we obtain~\cite{Supple},
\begin{align}
    [\mathcal{H}_{Q}^{0}]_{k,k^\prime} &= \delta_{ k, k^\prime}[\epsilon_\alpha(k+Q) - \epsilon_\alpha(k)+E_{\rm FM}],\\
    [\mathcal{H}_{Q}^{\rm int}]_{k,k^\prime} &= \sum_{\nu}\dfrac{U_\nu}{N}\Big\{\delta_{k, k^\prime}\sum_{q}\left[1-D_{\alpha,\alpha,\alpha,\alpha}^{\nu:Q}(k,q,q,k)\right]\notag\\
    &-\left[1-D_{\alpha,\alpha,\alpha,\alpha}^{\nu:Q}(k,k,k^\prime,k^\prime)\right]\Big\}.~\label{eq:Hint}
\end{align}
Here, we introduce $U_\nu=U/N_\nu$ with $N_\nu$ defined below. 
While $\mathcal{H}_Q^{0}$ arises from the energy excitation of a noninteracting Hamiltonian due to the spin flip, i.e., $\epsilon_\alpha(k+Q) - \epsilon_\alpha(k)$,
quantum geometry enters $\mathcal{H}_Q^{\rm int}$ through $D_{n,m,p,q}^{\nu:Q}(k_1,k_2,k_3,k_4) = 1-\bra{u_n(k_1+Q)}\tau_\nu\ket{ u_m(k_2)}\bra{u_p(k_3)}\tau_\nu^\dagger\ket{ u_q(k_4+Q)}$ with the diagonal matrices $\tau_\nu$, resembling the quantum distance $D_{nm}(k+Q,k) = 1-\vert\braket{u_n(k+Q)\vert u_m(k)}\vert^2$. 
$N_\nu$ and $\tau_\nu$ are chosen to satisfy $\sum_{\nu}[\tau_{\nu}]_{ii}[\tau_{\nu}^\dagger]_{jj}/N_\nu=\delta_{ij}$, and the following results do not depend on their choice.
From this formula, we find that the effective Coulomb interaction between Bloch electrons is tuned by quantum geometry through $D_{n,m,p,q}^{\nu:Q}(k_1,k_2,k_3,k_4)$, resulting in negative $E_{\rm spin}^{0}(Q)$. Thus, we conclude the quantum-geometry-driven magnetic phase transition.

\begin{figure}[tbp]
  \includegraphics[width=1.0\linewidth]{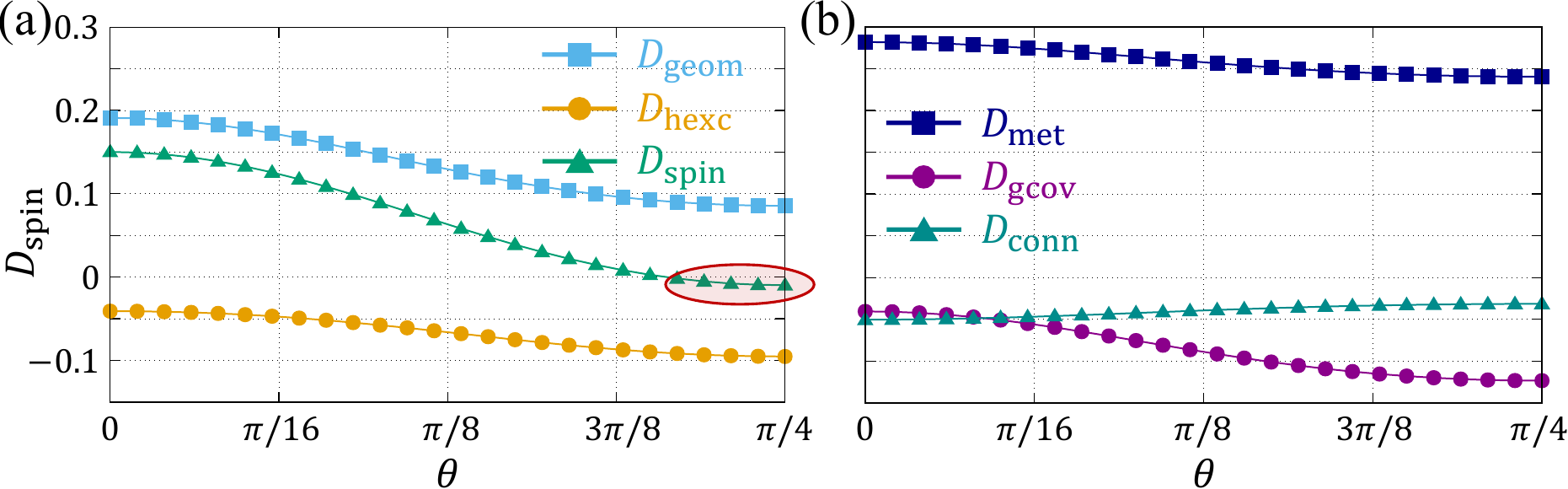}
  \centering
  \caption{$\theta$-dependence of the spin stiffness for $U = 1.36$. Panel (a) shows $\mathcal{D}_{\rm geom}$ (blue squares), $\mathcal{D}_{\rm hexc}$ (orange circles), $\mathcal{D}_{\rm spin }$ (green triangles), while panel (b) shows $\mathcal{D}_{\rm met}$ (dark-blue squares), $\mathcal{D}_{\rm \rm gcov}$ (dark-magenta circles), $\mathcal{D}_{\rm con }$ (dark-cyan triangles). In (a), the region highlighted by the red oval indicates the negative $\mathcal{D}_{\rm spin}$.}
  \label{fig:stiffness}
\end{figure}

Next, we derive the general expression of the spin stiffness~\cite{Supple}.
The spin stiffness is decomposed into the potentially positive quantum geometric contribution, $\mathcal{D}_{\rm geom}$, and the negative high-energy-excitation contribution, $\mathcal{D}_{\rm hexc}$, arising from the Stoner excitation and the optical magnons, as $\mathcal{D}_{\rm spin} = \mathcal{D}_{\rm geom} + \mathcal{D}_{\rm hexc}$.
In our model, the quantum geometric contribution 
consists of the quantum metric term $\mathcal{D}_{\rm met}$, the gauge-covariant derivative term $\mathcal{D}_{\rm \rm gcov}$, and the Berry connection term $\mathcal{D}_{\rm \rm conn}$, i.e., $\mathcal{D}_{\rm geom} = \mathcal{D}_{\rm met} + \mathcal{D}_{\rm \rm gcov} + \mathcal{D}_{\rm conn}$, where we define
\begin{align}
    \mathcal{D}_{\rm met}&= 2\sum_{\nu}\sum_{k,k^\prime}\dfrac{U_\nu}{N^2}g_{\alpha}(k)\tau_{\nu:{\alpha\alpha}}(k)[\tau_{\nu:{\alpha\alpha}}(k^\prime)]^*,
    \\
    \mathcal{D}_{\rm \rm gcov} &= -\sum_{\nu}\sum_{k,k^\prime}
    \sum_{n(\neq{\alpha})}\dfrac{U_\nu}{N^2}R_{{\alpha}n}(k)\tau_{\nu:n{\alpha}}(k)[\tau_{\nu:{\alpha\alpha}}(k^\prime)]^*\notag\\
    &+ {\rm c.c.},\label{eq:shift}\\
    \mathcal{D}_{\rm conn} &= -2\sum_{\nu}\dfrac{U_\nu}{N^2}\vert\sum_{k}\sum_{n(\neq{\alpha})}A_{{\alpha}n}(k)\tau_{\nu:n{\alpha}}(k)\vert^2,
\end{align}
with $\tau_{\nu:nm}(k) = \bra{u_n(k)}\tau_{\nu}\ket{u_m(k)}$.
This expression includes three well-known quantum geometric quantities: the Berry connection $A_{nm}(k) = -i\braket{u_n(k)\vert \partial_{k} u_m(k)}$, the quantum metric, $g_{n}(k) = \sum_{m(\neq n)}A_{nm}(k)A_{mn}(k)$, and the gauge covariant derivative of the Berry connection, $R_{nm}(k) =-i\mathscr{D}[A_{nm}(k)]$, with the gauge covariant derivative, $\mathscr{D}[\mathcal{O}] = \partial_{k}\mathcal{O}+i(A_{nn}(k)-A_{mm}(k))\mathcal{O}$.
See Appendix C for the full expression of $\mathcal{D}_{\rm spin}$, derived in a general setup.

Our complete formulation naturally disentangles the positive contribution of the quantum metric, favoring ferromagnetism, from other effects (see Appendix C for the proof of positivity). This differs from the previous expressions under the single-mode approximation for moiré systems, including the quantum metric~\cite{Wu2020}, and from those under the mean-field approximation, where the spin stiffness is generally not expressed in terms of the quantum metric~\cite{Kang2024}.
In particular, previous studies~\cite{Wu2020,Kang2024} did not account for the contribution from $\mathcal{D}_{\rm gcov}, \mathcal{D}_{\rm conn}$, and $\mathcal{D}_{\rm hexc}$, which are essential for the negative $\mathcal{D}_{\rm spin}$ as shown below.

We show the $\theta$ dependence of spin stiffness for $U = 1.36$ in Fig.~\ref{fig:stiffness} (a) and (b). In Fig.~\ref{fig:stiffness} (a), $\mathcal{D}_{\rm geom}$ (blue squares) is always positive, which is analytically proven in Appendix C.
However, near $\theta=\pi/4$ (highlighted by the red oval), we find that negative $\mathcal{D}_{\rm hexc}$ (orange circles) overcomes the positive $\mathcal{D}_{\rm geom}$, leading to the negative $\mathcal{D}_{\rm spin}$ (green triangles); saturated ferromagnetism is globally and locally unstable.
Moreover, in Fig.~\ref{fig:stiffness} (b), only the quantum-metric term $\mathcal{D}_{\rm met}$ (dark blue squares) is positive. This indicates that, in our model, the quantum metric plays a key role in stabilizing ferromagnetism; for the stability of ferromagnetism, the quantum metric must be large enough to overcome the other negative contributions.

This is consistent with the intuitive understanding of saturated ferromagnetism: the quantum metric is the gauge-invariant part of the Wannier-function spread~\cite{Marzari1997,Vanderbilt2018}, giving rise to long-range effective Coulomb interactions due to the Wannier-function overlaps. This yields the effective ferromagnetic exchange coupling, resulting in saturated ferromagnetism~\cite{Tasaki2020,Kitamura2025}. Thus, the quantum metric is the origin of saturated ferromagnetism~\cite{Tasaki2020,Kitamura2025}. In our study, this physical picture is explicitly justified by the microscopic calculation.

\textit{Conclusion---}In this Letter, we present a Hubbard model with a nearly flat band where tuning quantum geometry alone induces a magnetic phase transition and the quantum metric stabilizes ferromagnetism. Our study is based on the rigorous theory~\cite{Tasaki1995,Tanaka2001,Tanaka2003,Ueda2004,Lu2009,Tanaka2018,Tamura2019} and ferromagnetic spin-wave theory for the many-body Hamiltonian~\cite{Kusakabe1994,Tasaki1994,Tasaki1996,Penc1996,Neupert2012,Wu2020, Kang2024}, and is interpreted by the concept of quantum geometry. Therefore, our findings highlight the essential role of quantum geometry in the quantum many-body physics of dispersive band systems.

Our construction of a model with tunable quantum geometry and saturated ferromagnetism in quasi-1D systems can be naturally extended to higher dimensions~\cite{Supple}. The quantum geometric contribution to the spin stiffness, especially the positive quantum metric contribution, is not specific to a particular model (see Appendix C), which indicates that the quantum geometric origin of saturated ferromagnetism is broadly applicable. Thus, we expect that our study will stimulate future research on the interplay between quantum geometry and quantum many-body physics in dispersive-band systems beyond the mean-field framework.

\begin{acknowledgments}
We are grateful to C.-g. Oh, Y. Yanase and A. Daido for fruit discussion.
This work was supported by JSPS KAKENHI (Grant Nos. JP23KJ0783, JP25K23367, 
JP23K25783, JP23K25790, 
JP25H01246, and JP25H01252) and by the RIKEN TRIP initiative (RIKEN Quantum, Advanced General Intelligence for Science Program, Many-body Electron Systems).
\end{acknowledgments}

\textit{Note added---}While finalizing this manuscript, we became aware of a related preprint by J.~K.~Ding and M.~Claassen~\cite{ding2026quantum}.

\clearpage

\textit{Appendix A: Stoner magnetism---}The stoner factor is defined by $U\bar{\chi}(q)$ with momentum $q$ where $\bar{\chi}(q)$ is the maximum eigenvalue of the spin susceptibility matrix, $\chi(q)$.
The matrix element of the spin susceptibility is given by 
\begin{align}
    \chi_{ij}(q) = -\dfrac{T}{N}\sum_{k,\omega_n}\mathcal{G}_{ij}(\omega_n,k+q)\mathcal{G}_{ji}(\omega_n,k)
\end{align}
where $T$ is the temperature and $\mathcal{G}(\omega_n,k) = [i\omega_n-H(k)]^{-1}$ is the Green function with fermionic Matsubara freequency $\omega_n$.
When the Stoner criterion $U\bar{\chi}(q) \geq1$ is satisfied, the magnetic order corresponding to $\bar{\chi}(q)$ becomes stable~\cite{Supple}.

\begin{figure}[bp]
  \includegraphics[width=1.0\linewidth]{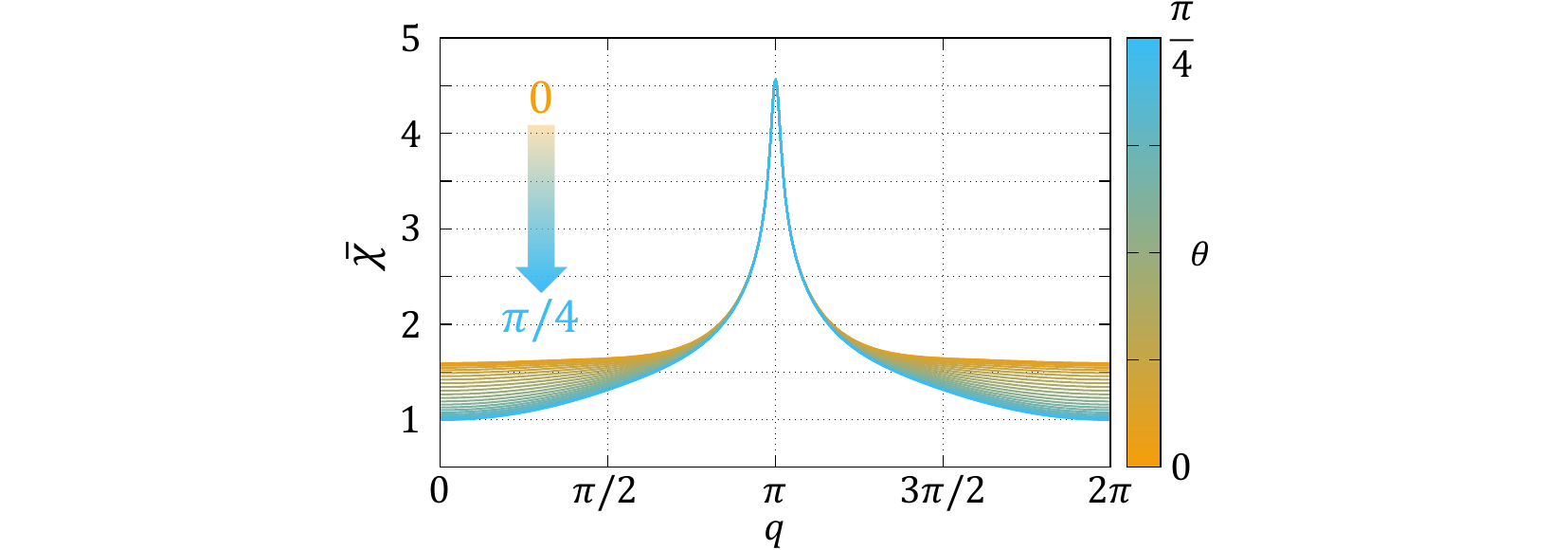}
  \centering
  \caption{(a) $\bm q$ dependence of  $\bar{\chi}(\bm q)$ for several values of $\theta$. The blue and orange lines correspond to $\theta = 0$ and $\pi/4$, respectively.}
  \label{fig:stoner_value}
\end{figure}

For the parameters analyzed in the main text, we show the $q$ dependence of $\bar{\chi}(q)$ in Fig.~\ref{fig:stoner_value}, where we set $T = 0.001$. Here, different colors indicate different values of $\theta$. For all $\theta$, we find that the maximum value of $\bar{\chi}(q)$ occurs at $q = \pi$, indicating the antiferromagnetic order. Thus, the Stoner theory does not reproduce our rigorous result. A detailed analysis is provided in the supplemental materials~\cite{Supple}.

\textit{Appendix B: Proof that $\varepsilon_0 = -s(\lambda^2+2)$ in the limit $t,v,U\rightarrow\infty$---}We analytically prove $\varepsilon_0 = -s(\lambda^2+2)$ in the limit of $t,v,U\rightarrow\infty$ for the local Hamiltonian,
\begin{align}
    &\hat{h}_{\bm x} = 
    \sum_{\sigma}\left[
    -s\hat{\alpha}_{\bm x,\sigma}^\dagger\hat{\alpha}_{\bm x,\sigma}
    +\dfrac{t^\prime}{2}\left(\hat{\beta}_{\bm x,\sigma}^\dagger\hat{\beta}_{\bm x,\sigma}
    +\hat{\beta}_{\bm x-\bm e,\sigma}^\dagger\hat{\beta}_{\bm x-\bm e,\sigma}
    \right)
    \right]
    \notag\\
    &+\sum_{\sigma}\sum_{i(\neq0,1)}
    \left[
    \dfrac{v^\prime}{2}\left(\hat{\gamma}_{i,\bm x,\sigma}^\dagger\hat{\gamma}_{i,\bm x,\sigma}+\hat{\gamma}_{i,\bm x-\bm e,\sigma}^\dagger\hat{\gamma}_{i,\bm x-\bm e,\sigma}\right)
    \right.
    \notag\\
    &+
    \left.
    u^\prime(\hat{\gamma}_{i,\bm x,\sigma}^\dagger\hat{\gamma}_{i,\bm x-\bm e,\sigma}+h.c.)
    \right]
    +(1-\kappa)U^\prime\hat{n}_{\bm x,\uparrow}\hat{n}_{\bm x,\downarrow}
    \notag\\
    &+
    \kappa\dfrac{U^\prime}{2}(
    \hat{n}_{\bm x+\bm e,\uparrow}\hat{n}_{\bm x+\bm e,\downarrow}
    +\hat{n}_{\bm x-\bm e,\uparrow}\hat{n}_{\bm x-\bm e,\downarrow}
    )
    \notag\\
    &+
    \dfrac{U^\prime}{2}\sum_{i\neq 0}\left(\hat{n}_{\bm x+\bm \mu_i,\uparrow}\hat{n}_{\bm x+ \bm \mu_i,\downarrow}
    +\hat{n}_{\bm x+\bm \mu_i-\bm e,\uparrow}\hat{n}_{\bm x+ \bm \mu_i-\bm e,\downarrow}
    \right),
\end{align}
with $t^\prime = t-\xi$, $v^\prime = v -\xi$, $u^\prime = u -\xi$, $U^\prime = U-\xi$.
Here, we restrict our attention to $0 < \kappa < 1$ while we adopt the case of $\kappa = 0$ for the numerical diagonalization in the main text.

\begin{figure}[tbp]
  \includegraphics[width=1.0\linewidth]{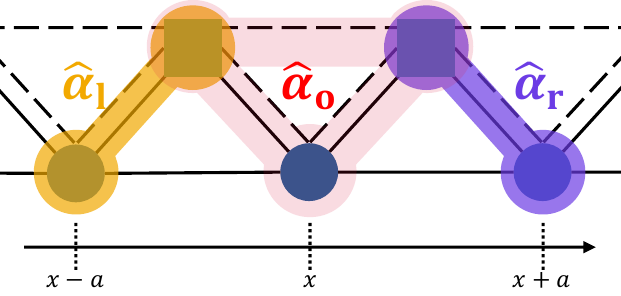}
  \centering
  \caption{
  Schematic illustration of $\hat{\alpha}^\dagger_{{\rm o},\sigma},\hat{\alpha}^\dagger_{{\rm r},\sigma}$, and $\hat{\alpha}^\dagger_{{\rm l},\sigma}$.
  }
  \label{fig:single_ele_basis}
\end{figure}

From the variational principle, we shall prove that $\lim_{t,v,U\uparrow\infty}\bra{\hat{\Phi}}\hat{h}_{\bm x}\ket{\hat{\Phi}}\geq-s(\lambda^2+2)$ for any normalized state $\ket{\hat{\Phi}}$, following Refs.~\cite{Tasaki1995,Tamura2019,Tasaki2020}.
We start by introducing a set of $(2N_{\rm sub}+1)$ operators that generate single-electron basis states for $\hat{h}_{\bm x}$.
Among them, $(2N_{\rm sub}-1)$ operators are chosen to be $\hat{\alpha}_{\bm x,\sigma}^\dagger,\hat{\beta}_{\bm x,\sigma}^\dagger\pm\hat{\beta}_{\bm x-\bm e,\sigma}^\dagger$, and $\hat{\gamma}_{i,\bm x,\sigma}^\dagger\pm\hat{\gamma}_{i,\bm x-\bm e,\sigma}^\dagger$ ($i\neq0,1$); the corresponding states are mutually orthogonal eigenstates of $\hat{h}_{\bm x}$. 
The remaining two operators are defined by $\hat{\alpha}_{{\rm l},\sigma}^\dagger = \lambda\hat{\gamma}_{0,\bm x-\bm e,\sigma}^\dagger-\hat{\gamma}_{1,\bm x-\bm e,\sigma}^\dagger$ and $\hat{\alpha}_{{\rm r},\sigma}^\dagger = \lambda\hat{\gamma}_{0,\bm x+\bm e,\sigma}^\dagger-\hat{\gamma}_{1,\bm x,\sigma}^\dagger$ (see Fig.~\ref{fig:single_ele_basis}).

Clearly, it suffices to consider the state such that $\lim_{t,v,U\uparrow\infty}\bra{\hat{\Phi}}\hat{h}_{\bm x}\ket{\hat{\Phi}}<\infty$, which we call the finite-energy condition. 
From the nonnegativity of $\hat{\beta}_{\bm x,\sigma}^\dagger\hat{\beta}_{\bm x,\sigma}$, $\hat{\gamma}_{i,\bm x,\sigma}^\dagger\hat{\gamma}_{i,\bm x,\sigma}$ $(i\neq 0,1)$, and $\hat{n}_{\bm r,\uparrow}\hat{n}_{\bm r,\downarrow}$, finite-energy condition reads $\hat{\beta}_{\bm x,\sigma}^\dagger\hat{\beta}_{\bm x,\sigma}\ket{\hat{\Phi}}=0$, $\hat{\gamma}_{i,\bm x,\sigma}^\dagger\hat{\gamma}_{i,\bm x,\sigma}\ket{\hat{\Phi}}=0$ ($i\neq 0,1$), and $\hat{c}_{\bm r,\uparrow}\hat{c}_{\bm r,\downarrow}\ket{\hat{\Phi}}=0$.
These conditions impose constraints on $\ket{\hat{\Phi}}$.

From the first and second conditions, it is easily found that $\ket{\hat{\Phi}}$ contains only $\{\hat{\alpha}^\dagger_{{\rm l},\sigma},\hat{\alpha}^\dagger_{{\rm o},\sigma},\hat{\alpha}^\dagger_{{\rm r},\sigma}\}$ that anticommute with $\hat{\beta}_{\bm x,\sigma}$ and $\hat{\gamma}_{i(\neq 0,1),\bm x,\sigma}$, where we define $\hat{\alpha}^\dagger_{{\rm o}, \bm x,\sigma}\equiv\hat{\alpha}^\dagger_{\bm x,\sigma}$.
In addition, $\hat{c}_{\bm x-\bm e,\sigma}^\dagger$, $\hat{c}_{\bm x,\sigma}^\dagger$, and $\hat{c}_{\bm x+\bm e,\sigma}^\dagger$ appear only in $\hat{\alpha}^\dagger_{{\rm l},\sigma}$, $\hat{\alpha}^\dagger_{{\rm o},\sigma}$, and $\hat{\alpha}^\dagger_{{\rm r},\sigma}$, respectively. Therefore, from $\hat{c}_{\bm r,\uparrow}\hat{c}_{\bm r,\downarrow}\ket{\hat{\Phi}}=0$ at $\bm r = \bm x,\bm x\pm\bm e$, we find that any site $z\in\{\rm l,o,r\}$ in terms of the $\alpha$ operators is not doubly occupied.
Therefore, the electron number $n_{\rm e}$ allowed in $\ket{\hat{\Phi}}$ is less than or equal to $3$.
In the following~\cite{Math_detail}, we show $\lim_{t,v,U\uparrow\infty}\bra{\hat{\Phi}}\hat{h}_{\bm x}\ket{\hat{\Phi}}\geq-s(\lambda^2+2)$ for $n_{\rm e} = 1,2$ and $3$.

Let $n_{\rm e} = 3$. 
A normalized state is written as $\ket{\hat{\Phi}_{3}}
=\sum_{\bm \sigma} C(\bm \sigma)\Big(\prod_{z}\hat{\alpha}_{z,\sigma_{z}}^\dagger\Big)\ket{0},
$ where $\bm \sigma=(\sigma_{\rm l},\sigma_{\rm o},\sigma_{\rm r})$ is summed over all possible spin configurations on $\{z\}$ and $C(\bm \sigma)$ is the corresponding coefficient.
Since $\hat{\alpha}_{\rm l (r), \sigma}^\dagger$ overlaps with $\hat{\alpha}_{{\rm o},\sigma}^\dagger$, the no-double-occupancy constraint
$\hat{c}_{\bm r,\uparrow}\hat{c}_{\bm r,\downarrow}\ket{\hat{\Phi}}=0$ at $\bm r=\bm x+\bm\mu_i$ and $\bm x+\bm\mu_i-\bm e$ ($i\neq 0$) leads to the relations $C(\sigma_{\rm l},\uparrow,\downarrow)=C(\sigma_{\rm l},\downarrow,\uparrow)$ and $C(\uparrow,\downarrow,\sigma_{\rm r})=C(\downarrow,\uparrow,\sigma_{\rm r})$. By using these relations repeatedly, we see that $C(\bm \sigma) = C(\bm \sigma^\prime)$ whenever the number of up (and down) spins in $\bm \sigma$ and $\bm \sigma^\prime$ is the same.
Hence, within a space of fixed $S_{\rm tot}^{(3)}$ which is the eigenvalue of $\sum_{\bm r}\hat{\bm c}_{\bm r}^\dagger\sigma_3\hat{\bm c}_{\bm r}/2$ with $\sigma_3={\rm diag}(1,-1)$, the state satisfying finite-energy condition is unique.
It follows that any finite-energy state must be fully spin-polarized, i.e., $S_{\rm tot} = 3/2$. Then, we can focus on $S_{\rm tot}^{(3)} = 3/2$, in which the problem reduces to solving the single-particle eigenvalue problem.

The case of $n_{\rm e} = 1$ is also the single-particle problem.
A straightforward calculation shows that the lowest single-particle eigenvalue of $\hat{h}_{\bm x}$ is $-s(\lambda^2+2)$, which is nondegenerate up to the trivial spin degeneracy, while other eigenvalues are nonnegative.
Thus, when $n_{\rm e} = 1,3$, we have $\lim_{t,v,U\uparrow\infty}\bra{\hat{\Phi}}\hat{h}_{\bm x}\ket{\hat{\Phi}}\geq-s(\lambda^2+2)$.

The remaining case is $n_{\rm e} = 2$.
For the triplet sector, i.e. $S_{\rm tot} = 1$, the above argument applies in the same way, and we again have $\lim_{t,v,U\uparrow\infty}\bra{\hat{\Phi}}\hat{h}_{\bm x}\ket{\hat{\Phi}}\geq-s(\lambda^2+2)$.
When $\ket{\hat{\Phi}}$ contains $\hat{\alpha}_{{\rm o},\sigma}^\dagger$, by using the no-double-occupancy condition, only the triplet states are allowed as in the $n_{\rm e}=3$ case.
Hence, it suffices to consider the singlet state without $\hat{\alpha}_{{\rm o},\sigma}^\dagger$ component, i.e., $\ket{\hat{\Phi}_{\rm s}} = C_{\rm s}(\hat{\alpha}_{\rm r,\uparrow}^\dagger\hat{\alpha}_{\rm l,\downarrow}^\dagger - \hat{\alpha}_{\rm r,\downarrow}^\dagger\hat{\alpha}_{\rm l,\uparrow}^\dagger)\ket{0}$, with a normalization coefficient $C_{\rm s}$. 
One readily finds $\bra{\hat{\Phi}_{\rm s}}\hat{h}_{\bm x}\ket{\hat{\Phi}_{\rm s}}=0$.
Consequently,
$\lim_{t,v,U\to\infty}\bra{\hat{\Phi}}\hat{h}_{\bm x}\ket{\hat{\Phi}}\ge -s(\lambda^2+2)$
holds for any electron number, which establishes that the ground state of $\hat{H}$ exhibits saturated ferromagnetism.

\textit{Appendix C: Spin stiffness for saturated ferromagnetism in isolated-half-filled band---}We consider the Hubbard Hamiltonian in any dimension, $\hat{H}^{\rm iso} = \hat{H}_{0}^{\rm iso}+\hat{H}_{\rm int}$, where the noninteracting part $\hat{H}_{0}^{\rm iso}$ possesses an isolated band denoted by the index $\tilde{n}$.
We assume that band $\tilde{n}$ is half-filled and is isolated by large band gaps. 
Then, we define the saturated ferromagnetic state for the isolated band as, $\ket{\hat{\phi}_{\rm all\uparrow}^{\rm iso}}=\prod_{\bm k}\hat{d}_{\tilde{n},\uparrow}^\dagger(\bm k)\ket{\tilde{0}}$ where the wave vector ${\bm k}$ runs over the Brillouin zone. Here, $\ket{\tilde{0}}$ is a reference state satisfying $\braket{\tilde{0}\vert\tilde{0}}=1$ in which all bands below $\tilde{n}$ are completely filled.
The spin-excitation energy and spin stiffness for $\ket{\hat{\phi}_{\rm all\uparrow}^{\rm iso}}$ are defined in a similar manner to the main text by replacing $\ket{\hat{\phi}_{\rm all\uparrow}}, \alpha$, and $E_{\rm FM}$, with $\ket{\hat{\phi}_{\rm all\uparrow}^{\rm iso}}, \tilde{n}$, and $E_{\rm FM}^{\rm iso} = \bra{\hat{\phi}_{\rm all\uparrow}^{{\rm iso}}}\hat{H}^{\rm iso}\ket{\hat{\phi}_{\rm all\uparrow}^{{\rm iso}}}$, respectively.

In systems with more than one dimension, spin stiffness is defined as a symmetric matrix with respect to the spatial dimensions, and we can adopt a coordinate that diagonalizes this matrix. Therefore, it suffices to consider the diagonal component $\mathcal{D}^{\mu\mu}=\mathcal{D}_{\rm geom}^{\mu\mu}+\mathcal{D}_{\rm hexc}^{\mu\mu}$ where $\mu$ indicates a spatial coordinate, even in two and three dimensions.
First, the quantum geometric term $\mathcal{D}_{\rm geom}^{\mu\mu}$ is divided into three components as $\mathcal{D}_{\rm geom}^{\mu\mu} = \mathcal{D}_{\rm met}^{\mu\mu} + \mathcal{D}_{\rm \rm gcov}^{\mu\mu} + \mathcal{D}_{\rm conn}^{\mu\mu}$ with
\begin{align}
    \mathcal{D}_{\rm met}^{\mu\mu}&= 2\sum_{\nu}\sum_{\bm k,\bm k^\prime}\dfrac{U_\nu}{N^2}g_{\tilde{n}}^{\mu\mu}(\bm k)\tau_{\nu:{\tilde{n}\tilde{n}}}(\bm k)[\tau_{\nu:{\tilde{n}\tilde{n}}}(\bm k^\prime)]^*,
    \\
    \mathcal{D}_{\rm \rm gcov}^{\mu\mu} &=\sum_{\nu}\sum_{\bm k,\bm k^\prime}
    \sum_{n(\neq{\tilde{n}})}\dfrac{U_\nu}{N^2}[f_{\tilde{n}n}^{\mu\mu}(\bm k)-R_{{\tilde{n}}n}^{\mu\mu}(\bm k)]\notag\\
    &\times\tau_{\nu:n{\tilde{n}}}(\bm k)[\tau_{\nu:{\tilde{n}\tilde{n}}}(\bm k^\prime)]^*+{\rm c.c.},\\
    \mathcal{D}_{\rm conn}^{\mu\mu} &= -2\sum_{\nu}\dfrac{U_\nu}{N^2}\vert\sum_{\bm k}\sum_{n(\neq{\tilde{n}})}A_{{\tilde{n}}n}^{\mu}(\bm k)\tau_{\nu:n{\tilde{n}}}(\bm k)\vert^2.
\end{align}
Here, the quantum geometric quantities are defined
as
$A_{nm}^\mu(\bm k) = -i\braket{u_n(\bm k)\vert\partial_{k_\mu} u_m(\bm k)}$, $g_{n}^{\mu\mu}(\bm k) = \sum_{m(\neq n)}A_{nm}^\mu(\bm k)A_{mn}^\mu(\bm k)$, and $R_{nm}^{\mu\mu}(\bm k) =-i\mathscr{D}_\mu[A_{nm}^\mu(\bm k)]$ with $\mathscr{D}_\mu[\mathcal{O}] = \partial_{k_\mu}\mathcal{O}+i(A_{nn}^\mu(\bm k)-A_{mm}^\mu(\bm k))\mathcal{O}$
and $f_{nm}^{\mu\mu}(\bm k) = \sum_{p(\neq n,m)}A_{np}^\mu(\bm k)A_{pm}^\mu(\bm k)$, with $N$ being the number of unit cells.

The quantity $f_{nm}^{\mu\mu}(\bm k)$, resembling the quantum metric, requires two different non-zero Berry connections, $A_{np}^\mu(\bm k)$ and $A_{pm}^\mu(\bm k)$ with $n\neq m$, $m\neq p$, and $p\neq n$, to be finite.
However, in the model of the main text, only $A_{\alpha\beta}^\mu(\bm k)$ is non-zero where $\beta$ indicates the band corresponding to $\hat{\beta}_{\bm x,\sigma}^\dagger$ since the Bloch wave functions of the bands corresponding to $\hat{\gamma}_{i(\neq0,1),\bm x,\sigma}$ are momentum independent.
Therefore, $f_{nm}^{\mu\mu}(\bm k)$ vanishes and does not appear in Eq.~\eqref{eq:shift} of the main text.

Let us discuss the sign of each term.
From the relationship $\sum_{\nu}\tau_{\nu:{\tilde{n}\tilde{n}}}(\bm k)\tau_{\nu:{\tilde{n}\tilde{n}}}(\bm k^\prime)/N_\nu=\sum_{n}\vert u_{\tilde{n}:n}(\bm k)\vert^2
\vert u_{\tilde{n}:n}(\bm k^\prime)\vert^2$, where $u_{\tilde{n}:n}(\bm k)$ is the $n$th component of $\ket{ u_{\tilde{n}}(\bm k)}$, the quantum metric contribution is rewritten as 
$ 
    \mathcal{D}_{\rm met}^{\mu\mu}= \frac{2U}{N^2}\sum_{\bm k,\bm k^\prime}g_{\tilde{n}}^{\mu\mu}(\bm k)\sum_{n}\vert u_{\tilde{n}:n}(\bm k)\vert^2
\vert u_{\tilde{n}:n}(\bm k^\prime)\vert^2.    
$ 
Therefore, $g_{n}^{\mu\mu}(\bm k)\geq0$ leads to $\mathcal{D}_{\rm met}^{\mu\mu}\geq0$.
As for the other two terms, while $\mathcal{D}_{\rm conn}^{\mu\mu}$ is clearly non-positive, the sign of $\mathcal{D}_{\rm \rm gcov}^{\mu\mu}$ is uncertain.
As a result, the total sign of $\mathcal{D}_{\rm geom}^{\mu\mu}$ is uncertain.
However, we show $\mathcal{D}_{\rm geom}>0$ for the Hamiltonian in the main text.
We can show that the Bloch wave function of $\hat{H}_0$ is independent of the value of $s$. 
Therefore, $\mathcal{D}_{\rm geom}$ is determined only by the Bloch wave function, 
meaning that $\mathcal{D}_{\rm geom}$ of $\hat{H}$ is the same as that of $\hat{H}_{\rm flat}=\hat{H}\vert_{s=0}$.
Considering that the saturated ferromagnetic state is the unique ground state of $\hat{H}_{\rm flat}$, the spin stiffness of $\hat{H}_{\rm flat}$ should be positive, which indicates $\mathcal{D}_{\rm geom}>0$ for any $s$ (note that $\mathcal{D}_{\rm hexc}^{\mu\mu}$ is negative as shown below).

Next, the high-energy excitation contribution is given by 
\begin{align}
    &\mathcal{D}_{\rm hexc}^{\mu\mu}
    =-2\sum_{\chi > 0}\dfrac{\vert[\bm A_{\bm 0}^0]^\dagger\varDelta\mathcal{H}_\mu
    \bm A_{\bm 0}^{\chi}\vert^2}{E^\chi_{\rm spin}(\bm 0)},\notag\\
    &[\varDelta\mathcal{H}_\mu]_{\bm k,\bm k^\prime}
    \notag\\
    &= \delta_{\bm k,\bm k^\prime}\partial_{k_\mu}\Big[\epsilon_{\tilde{n}}(\bm k) 
    + \sum_{\nu}\sum_{n(\leq\tilde{n})}\sum_{\bm q}\dfrac{U_\nu}{N}\vert\tau_{\nu:n\tilde{n}}(\bm k,\bm q)\vert^2\Big]\notag\\
    &+i\sum_{\nu}\sum_{n(\neq{\tilde{n}})}\dfrac{U_\nu}{N}
    \Big(A^{\mu}_{\tilde{n}n}(\bm k)\tau_{\nu:n\tilde{n}}(\bm k)[\tau_{\nu:\tilde{n}\tilde{n}}(\bm k^\prime)]^*\notag\\
    &-\tau_{\nu:\tilde{n}\tilde{n}}(\bm k)[A^{\mu}_{\tilde{n}n}(\bm k^\prime)\tau_{\nu:n\tilde{n}}(\bm k^\prime)]^*
    \Big).
\end{align}
Here, $n<\tilde{n}$ indicates the bands below $\tilde{n}$.
Because of $E_{\rm spin}^{\chi(> 0)}(\bm 0)>0$, $\mathcal{D}_{\rm hexc}^{\mu\mu}$ is always negative. 
Note that, since $E_{\bm 0}^\chi$ and $\bm A_{\bm 0}^\chi$ are independent of the energy dispersion and are determined only by $\mathcal{H}_{\bm 0}^{\rm int}$, the energy dispersion contributes to the spin stiffness via only its group velocity $\partial_{k_\mu}\epsilon_n(\bm k)$.

\EnableTOC

\clearpage

\renewcommand{\thesection}{S\arabic{section}}
\renewcommand{\theequation}{S\arabic{equation}}
\setcounter{equation}{0}
\renewcommand{\thefigure}{S\arabic{figure}}
\setcounter{figure}{0}
\renewcommand{\thetable}{S\arabic{table}}
\setcounter{table}{0}
\makeatletter
\c@secnumdepth = 2
\makeatother

\onecolumngrid

\begin{center}
 {\large \textmd{Supplemental Materials:} \\[0.3em]
 {\bfseries Quantum-geometry-driven exact ferromagnetic ground state in a nearly-flat band}}
\end{center}

\setcounter{page}{1}

\tableofcontents
\section{Saturated ferromagnetism in isolated half-filled band}

We discuss the saturated ferromagnetism in an isolated half-filled band.
Although the main text focuses on a one-dimensional system and a specific model, the following discussion is general and applies to any spatial dimension ($d=1,2,3$).

\subsection{Saturated ferromagnetism}

We consider the Hubbard model with $SU(2)$ symmetry defined by
\begin{align}
    \hat{H} &= \hat{H}_0+\hat{H}_{\rm int},\\
    \hat{H}_0 &= \sum_{i,j}\sum_{\sigma}\sum_{\bm R,\bm R^\prime}\hat{c}^\dagger_{i,\sigma}(\bm R)t_{i,j}(\bm R-\bm R^\prime)\hat{c}_{j,\sigma}(\bm R^\prime),\\
    \hat{H}_{\rm int} &= U\sum_{i}\sum_{\bm R}
    \hat{c}_{i,\uparrow}^\dagger(\bm R)\hat{c}_{i,\uparrow}(\bm R)
    \hat{c}_{i,\downarrow}^\dagger(\bm R)\hat{c}_{i,\downarrow}(\bm R),
\end{align}
where $\hat{H}_0$ is the noninteracting Hamiltonian and $\hat{H}_{\rm int}$ represents the Coulomb interaction.
Here, $\hat{c}^\dagger_{i,\sigma}(\bm R)$ is the fermionic creation operator at position$\bm R = (R_1, \ldots, R_d)$ with spin $\sigma = \uparrow, \downarrow$ and an internal degree of freedom $i$ (e.g., orbital or sublattice). $t_{i,j}(\bm R-\bm R^\prime)$ is the spin-independent hopping integral between $(\bm R, i)$ and $(\bm R^\prime, j)$, and $U>0$ represents the on-site Coulomb interaction. 
We impose periodic boundary conditions and denote by $N$ the total number of unit cells, and by $f$ the number of internal degrees of freedom per unit cell (excluding spin).
The fermionic operators obey the anticommutation relations
\begin{align}
    \{\hat{c}^\dagger_{i,\sigma}(\bm R),\hat{c}_{j,\sigma^\prime}(\bm R^\prime)\} &= \delta_{i,j}\delta_{\sigma,\sigma^\prime}\delta_{\bm R,\bm R^\prime},\\
    \{\hat{c}^\dagger_{i,\sigma}(\bm R),\hat{c}_{j,\sigma^\prime}^\dagger(\bm R^\prime)\} &=0,\\
    \{\hat{c}_{i,\sigma}(\bm R),\hat{c}_{j,\sigma^\prime}(\bm R^\prime)\} &=0.
\end{align}

Then, we define the fermionic operators in momentum space. Due to the sublattice degrees of freedom, there are two ways to define the Fourier transformation: one that excludes the internal position of sublattices and one that includes it. In the latter definition, fermionic operators in momentum space are defined in the non-periodic form. This non-periodicity causes technical difficulties for the formulation. Thus, we adopt the former one, namely the Fourier transformation in the \textit{periodic} form, as
\begin{align}
    \hat{c}^\dagger_{i,\sigma}(\bm k) = \dfrac{1}{\sqrt{N}}\sum_{\bm R}e^{i\bm k\cdot \bm R} \hat{c}_{i,\sigma}^\dagger(\bm R),~\label{eq:periodic_Fourier}
\end{align}
where $\bm k$ is the vector of the crystal momentum 
The anticommutation relations for the fermionic operator in momentum space are given by
\begin{align}
    \{\hat{c}^\dagger_{i,\sigma}(\bm k),\hat{c}_{j,\sigma^\prime}(\bm k^\prime)\} 
    &= \delta_{i,j}\delta_{\sigma,\sigma^\prime}\delta_{\bm k-\bm k^\prime, 2\pi \bm n},\\
    \{\hat{c}^\dagger_{i,\sigma}(\bm k),\hat{c}_{j,\sigma^\prime}^\dagger(\bm k^\prime)\} &=0,\\
    \{\hat{c}_{i,\sigma}(\bm k),\hat{c}_{j,\sigma^\prime}(\bm k^\prime)\} &=0, 
\end{align}
where $\bm n = (n_1, \ldots, n_d)$ is the vector of integers $n_1,\ldots,n_d$.

Using this periodic convention in Eq.~\eqref{eq:periodic_Fourier}, the momentum-space representation of the Hubbard model is given by
\begin{align}
    \hat{H}_0 &= \sum_{\sigma}\sum_{\bm k}\hat{\bm c}_{\sigma}^\dagger(\bm k)H(\bm k)\hat{\bm c}_{\sigma}(\bm k)\\
    \hat{H}_{\rm int} &= \dfrac{U}{N}\sum_{i}\sum_{\bm k,\bm k^\prime,\bm q}\hat{c}_{i,\downarrow}^\dagger(\bm k+\bm q)\hat{c}_{i,\uparrow}^\dagger(-\bm k)
    \hat{c}_{i,\uparrow}(-\bm k^\prime)\hat{c}_{i,\downarrow}(\bm k^\prime+\bm q),
\end{align}
with the vector representation of the creation operators $\hat{\bm c}_{\sigma}^\dagger(\bm k) =(\hat{c}_{1,\sigma}^\dagger(\bm k), \cdots, \hat{c}_{f,\sigma}^\dagger(\bm k))$ and the matrix representation of the noninteracting Hamiltonian $H(\bm k)$ in momentum space.
The matrix component of $H(\bm k)$ is given by $[H(\bm k)]_{i,j} = \sum_{ \bm r}e^{-i\bm k\cdot\bm r}t_{ij}(\bm r)$ with $\bm r = \bm R - \bm R^\prime$.
The noninteracting Hamiltonian is diagonalized as
\begin{align}
    H(\bm k)\ket{u_n(\bm k)} = \epsilon_n(\bm k)\ket{u_n(\bm k)},
\end{align}
with the $2\pi$-periodic part of the Bloch-wave function $\ket{u_n(\bm k)}$ satisfying $\braket{u_m(\bm k)\vert u_n(\bm k)}=\delta_{n,m}$ and the energy dispersion $\epsilon_n(\bm k)$ of the $n$th band.
We denote the single particle eigenstate of $\hat{H}_0$ with eigenvalue $\epsilon_n(\bm k)$ as $\hat{d}_{n,\sigma}^\dagger(\bm k)\ket{0}$, where $\ket{0}$ is the vacuum state.

Hereafter, we assume that the band labeled by $\tilde{n}$ is isolated from the other bands and that the band gaps above and below it are large enough to exceed $U$.
In this setup, the saturated ferromagnetic state of the half-filled isolated band is defined by
\begin{align}
    \ket{\hat{\psi}_{\rm all \uparrow}} = \prod_{\bm k}\hat{d}_{\tilde{n},\uparrow}^\dagger(\bm k)\ket{\tilde{0}}.
\end{align}
Here, $\ket{\tilde{0}}$ is the normalized reference state in which all bands below $\tilde{n}$ are filled. 
The total energy of $\ket{\hat{\psi}_{\rm all \uparrow}}$ is defined by
\begin{align}
    E_{\rm FM} &= \bra{\hat{\psi}_{\rm all \uparrow}}\hat{H}\ket{\hat{\psi}_{\rm all \uparrow}}.
\end{align}

\subsection{Spin excitation energy}

We define the one-spin-flipped state from the saturated ferromagnetic state within the isolated band as
\begin{align}
  \ket{\hat{\phi}_{\bm Q;\bm k}} = \hat{d}_{\tilde{n},\downarrow}^\dagger(\bm k+\bm Q)
  \hat{d}_{\tilde{n},\uparrow}(\bm k)\ket{\hat{\psi}_{\rm all\uparrow}},
\end{align}
with momentum $\bm Q$.
Because of the band gaps, the low-lying excitation energy above the saturated ferromagnetic state, which we call the spin excitation energy, is expected to be well approximated by the effective Hamiltonian in the subspace spanned by $\ket{\hat{\phi}_{\bm Q;\bm k}}$~\cite{SKusakabe1994,STasaki1994,STasaki1996,SPenc1996,SNeupert2012}.
Since $\bm Q$ is a good quantum number, the effective Hamiltonian is labeled by $\bm Q$ as $\mathcal{H}_{\bm Q}$, whose matrix elements are given by $[\mathcal{H}_{\bm Q}]_{\bm k,\bm k^\prime} = \bra{\hat{\phi}_{\bm Q;\bm k}}\hat{\mathcal{H}}\ket{\hat{\phi}_{\bm Q;\bm k^\prime}}$.
We write the eigenvalue equationfor $\mathcal{H}_{\bm Q}$ as
\begin{align}
    \mathcal{H}_{\bm Q}\bm A_{\bm Q}^{\chi} = E_{\bm Q}^{\chi}\bm A_{\bm Q}^{\chi},
\end{align}
with the index of the eigenvalue, $\chi$.
We also write the vector component of $\bm A_{\bm Q}^{\chi}$ by $A_{\bm Q}^{\chi}(\bm k)$.
Note that $E^{\chi}_{\bm Q}$ is also given by
$
    E^{\chi}_{\bm Q} = \bra{\hat{\phi}_{\bm Q}^{\chi}}\hat{\mathcal{H}}\ket{\hat{\phi}_{\bm Q}^{\chi}}
$
with
$    
    \ket{\hat{\phi}_{\bm Q}^\chi} = \sum_{\bm k}A_{\bm Q}^\chi(\bm k)\ket{\hat{\phi}_{\bm Q;\bm k}}.
$
Then, the spin excitation energy of the saturated ferromagnetism is defined as~\cite{SKusakabe1994,STasaki1994,STasaki1996,SPenc1996,SNeupert2012,SWu2020, SKang2024}
\begin{align}
    E_{\rm spin}^\chi(\bm Q) = \bra{\hat{\phi}_{\bm Q}^\chi}\hat{\mathcal{H}}\ket{\hat{\phi}_{\bm Q}^\chi} - E_{\rm FM}.
\end{align}
For the calculation, it is useful to divide $\bra{\hat{\phi}_{\bm Q}^\chi}\hat{\mathcal{H}}\ket{\hat{\phi}_{\bm Q}^\chi}$  into two parts as
\begin{align}
   \mathcal{H}_{\bm Q} &=\mathcal{H}^{0}_{\bm Q} + \mathcal{H}^{\rm int}_{\bm Q},\\
   [\mathcal{H}_{\bm Q}^{0}]_{\bm k,\bm k^\prime} &= \bra{\hat{\phi}_{\bm Q;\bm k}}\hat{H}_0\ket{\hat{\phi}_{\bm Q;\bm k^\prime}},\\
   [\mathcal{H}_{\bm Q}^{\rm int}]_{\bm k,\bm k^\prime} &= \bra{\hat{\phi}_{\bm Q;\bm k}}\hat{H}_{\rm int}\ket{\hat{\phi}_{\bm Q;\bm k^\prime}}.
\end{align}
In the following, we derive the Bloch representation of $\mathcal{H}_{\bm Q}^{0}$ and $\mathcal{H}_{\bm Q}^{\rm int}$.

\subsubsection{Noninteracting term}

We introduce the Bloch representation of the noninteracting Hamiltonian as
\begin{align}
    \hat{H}_0 &= \sum_{\sigma}\sum_{n}\sum_{\bm k}\epsilon_n(\bm k)\hat{d}_{n,\sigma}^\dagger(\bm k)\hat{d}_{n,\sigma}(\bm k)~\label{eq:H0_Bloch}.
\end{align}
By acting with $\epsilon_n(\bm k)\hat{d}_{n,\uparrow}^\dagger(\bm k)\hat{d}_{n,\uparrow}(\bm k)$ and $\epsilon_n(\bm k)\hat{d}_{n,\downarrow}^\dagger(\bm k)\hat{d}_{n,\downarrow}(\bm k)$ on $\ket{\hat{\phi}_{\bm Q;\bm k}}$, we obtain
\begin{align}
    \epsilon_n(\bm k)\hat{d}_{n,\uparrow}^\dagger(\bm k)\hat{d}_{n,\uparrow}(\bm k)\ket{\hat{\phi}_{\bm Q;\bm k^\prime}}
    &=
    \epsilon_{n}(\bm k)\sum_{m(\leq\tilde{n})}\delta_{n,m}\left(1-\hat{d}_{m,\uparrow}(\bm k)\hat{d}_{m,\uparrow}^\dagger(\bm k)\right)\hat{d}_{\tilde{n},\downarrow}^\dagger(\bm k^\prime+\bm Q)
    \hat{d}_{\tilde{n},\uparrow}(\bm k^\prime)\ket{\hat{\psi}_{\rm all\uparrow}}\notag\\
    &=\epsilon_{n}(\bm k)\Big[\sum_{m(\leq\tilde{n})}\delta_{n,m}
    \ket{\hat{\phi}_{\bm Q;\bm k^\prime}}\notag\\
    &-\delta_{n,\tilde{n}}
    \hat{d}_{\tilde{n},\downarrow}^\dagger(\bm k^\prime+\bm Q)\hat{d}_{\tilde{n},\uparrow}(\bm k)
    \left(
    \delta_{\bm k,\bm k^\prime}-\hat{d}_{\tilde{n},\uparrow}(\bm k^\prime)\hat{d}_{\tilde{n},\uparrow}^\dagger(\bm k)
    \right)\ket{\hat{\psi}_{\rm all\uparrow}}\Big]\notag\\
    &=
    \epsilon_{n}(\bm k)\Big[
    \sum_{m(\leq \tilde{n})}\delta_{n,m}-\delta_{n,\tilde{n}}\delta_{\bm k,\bm k^\prime}
    \Big]
    \ket{\hat{\phi}_{\bm Q;\bm k^\prime}},\\
    \epsilon_n(\bm k)\hat{d}_{n,\downarrow}^\dagger(\bm k)\hat{d}_{n,\downarrow}(\bm k)\ket{\hat{\phi}_{\bm Q;\bm k^\prime}}
    &=
    \epsilon_{n}(\bm k)\Big[
    \sum_{m(<\tilde{n})}\delta_{n,m}\ket{\hat{\phi}_{\bm Q;\bm k^\prime}}
    \notag\\
    &+\delta_{n,\tilde{n}}\hat{d}_{\tilde{n},\downarrow}^\dagger(\bm k)\left(\delta_{\bm k,\bm k^\prime+\bm Q}-\hat{d}_{\tilde{n},\downarrow}^\dagger(\bm k^\prime+\bm Q)\hat{d}_{\tilde{n},\downarrow}(\bm k)\right)
    \hat{d}_{\tilde{n},\uparrow}(\bm k^\prime)\ket{\hat{\psi}_{\rm all\uparrow}}
    \Big]
    \notag\\
    &=
    \epsilon_{n}(\bm k)
    \Big[
    \sum_{m(<\tilde{n})}\delta_{n,m}
    +\delta_{n,\tilde{n}}\delta_{\bm k,\bm k^\prime+\bm Q}
    \Big]
    \ket{\hat{\phi}_{\bm Q;\bm k^\prime}},
\end{align}
where we use $\hat{d}_{n,\sigma}^\dagger(\bm k)\hat{d}_{n,\sigma}^\dagger(\bm k) = 0$ and $\hat{d}_{n,\sigma}(\bm k^\prime)\ket{0} = 0$.
By using these relations, we obtain,
\begin{align}
    [\mathcal{H}_{\bm Q}^{0}]_{\bm k,\bm k^\prime} = 
    \delta_{\bm k,\bm k^\prime}\Big[\epsilon_{\tilde{n}}(\bm k+\bm Q)- \epsilon_{\tilde{n}}(\bm k) +\bra{\hat{\psi}_{\rm all\uparrow}}\hat{H}_0\ket{\hat{\psi}_{\rm all\uparrow}}\Big].
\end{align}

\subsubsection{Coulomb interaction term}

The starting point is the decomposition of the Coulomb interaction,
\begin{align}
    \hat{H}_{\rm int} &= \dfrac{U}{N}\sum_{i}\sum_{\bm k,\bm k^\prime,\bm q}\hat{c}_{i,\downarrow}^\dagger(\bm k+\bm q)\hat{c}_{i,\uparrow}^\dagger(-\bm k)
    \hat{c}_{i,\uparrow}(-\bm k^\prime)\hat{c}_{i,\downarrow}(\bm k^\prime+\bm q)\notag\\
    &=-\dfrac{U}{N}\sum_{i}\sum_{\bm k,\bm k^\prime,\bm q}
    \left[
    \hat{c}_{i,\downarrow}^\dagger(\bm k+\bm q)\hat{c}_{i,\uparrow}(-\bm k^\prime)\hat{c}_{i,\uparrow}^\dagger(-\bm k)
    \hat{c}_{i,\downarrow}(\bm k^\prime+\bm q)
    -\delta_{\bm k,\bm k^\prime}\hat{c}_{i,\downarrow}^\dagger(\bm k+\bm q)
    \hat{c}_{i,\downarrow}(\bm k+\bm q)
    \right]\notag\\
    &=-\sum_{\bm k,\bm k^\prime,\bm q}    
    \left[
    \sum_{\nu}\dfrac{U_\nu}{N}\hat{\bm c}_{\downarrow}^\dagger(\bm k+\bm q)\tau_\nu\hat{\bm c}_{\uparrow}(\bm k)\hat{\bm c}_{\uparrow}^\dagger(\bm k^\prime)
    \tau_\nu^\dagger\hat{\bm c}_{\downarrow}(\bm k^\prime+\bm q)
    -\dfrac{U}{N}\delta_{\bm k,\bm k^\prime}\hat{\bm c}_{\downarrow}^\dagger(\bm k+\bm q)
    \hat{\bm c}_{\downarrow}(\bm k+\bm q)
    \right].~\label{eq:Hint_multi}
\end{align}
In the third line, we use the variable transformation~\cite{variable}.
Here, we introduce $U_\nu = U/N_\nu$ and the diagonal matrices $\tau_\nu$. $N_\nu$ and  $\tau_\nu$ are determined to satisfy $\sum_{\nu}[\tau_\nu]_{i,i}[\tau_\nu^\dagger]_{j,j}/N_\nu = \delta_{i,j}$.
One choice of them is to choose the set of matrices $\{\tau_\mu\}$ such that the vectors formed by their diagonal entries constitute an orthogonal basis of the $f$-dimensional vector space.
More precisely, for each $\tau_\mu$ we define an $f$-component vector $\bm d_\mu \equiv \bigl((\tau_\mu)_{11},(\tau_\mu)_{22},\ldots,(\tau_\mu)_{ff}\bigr)^{\mathsf T}$, and require that $\{\bm d_\mu\}$ form an orthogonal basis of the $f$-dimensional vector space.
With this choice, the matrices $\{\tau_\mu\}$ together with $N_\tau={\rm Tr}[\tau_\mu^\dagger\tau_\mu]$ satisfy the above condition.
In the following, we do not specify $N_\nu$ and $\tau_{\nu}$ since the following results are independent of the choice of $\tau_\nu$ and $N_\nu$.

By inserting the identity matrix $\bm 1=\sum_{n}\ket{u_n(\bm k)}\bra{u_n(\bm k)}$ into both the right side of each creation operator and the left side of each annihilation operator of Eq.~\eqref{eq:Hint_multi}, we obtain
\begin{align}
    \hat{H}_{\rm int}&=-\sum_{\nu}\sum_{nmpq}\sum_{\bm k,\bm k^\prime,\bm q}
    \dfrac{U_\nu}{N}
    \tau_{\nu:nm}^{\bm q}(\bm k,\bm k)
    [\tau_{\nu:qp}^{\bm q}(\bm k^\prime,\bm k^\prime)]^*
    \hat{d}_{n,\downarrow}^\dagger(\bm k+\bm q)\hat{d}_{m,\uparrow}(\bm k)\hat{d}_{p,\uparrow}^\dagger(\bm k^\prime)
    \hat{d}_{q,\downarrow}(\bm k^\prime+\bm q)\notag\\
    &+\dfrac{U}{N}\sum_n\sum_{\bm k,\bm q}\hat{d}_{n,\downarrow}^\dagger(\bm k+\bm q)
    \hat{d}_{n,\downarrow}(\bm k+\bm q)\notag\\
    &=\sum_{\nu}\sum_{nmpq}\sum_{\bm k,\bm k^\prime,\bm q}
    \dfrac{U_\nu}{N}
    \tau_{\nu:nm}^{\bm q}(\bm k,\bm k)
    [\tau_{\nu:qp}^{\bm q}(\bm k^\prime,\bm k^\prime)]^*
    \hat{d}_{n,\downarrow}^\dagger(\bm k+\bm q)
    \hat{d}_{p,\uparrow}^\dagger(\bm k^\prime)
    \hat{d}_{m,\uparrow}(\bm k)
    \hat{d}_{q,\downarrow}(\bm k^\prime+\bm q)\notag\\
    &
    +\sum_{nmp}
    \sum_{\bm k,\bm q}\bigg\{
    \dfrac{U}{N}\delta_{n,m}\delta_{mp}
    -\sum_\nu\dfrac{U_\nu}{N}
    \tau_{\nu:nm}^{\bm q}(\bm k,\bm k)
    [\tau_{\nu:pm}^{\bm q}(\bm k,\bm k)]^*
    \bigg\}\hat{d}_{n,\downarrow}^\dagger(\bm k+\bm q)
    \hat{d}_{p,\downarrow}(\bm k+\bm q)\notag\\
    &=\sum_{\nu}\sum_{nmpq}\sum_{\bm k,\bm k^\prime,\bm q}
    \dfrac{U_\nu}{N}
    \tau_{\nu:nm}^{\bm q}(\bm k,\bm k)
    [\tau_{\nu:qp}^{\bm q}(\bm k^\prime,\bm k^\prime)]^*
    \hat{d}_{n,\downarrow}^\dagger(\bm k+\bm q)
    \hat{d}_{p,\uparrow}^\dagger(\bm k^\prime)
    \hat{d}_{m,\uparrow}(\bm k)
    \hat{d}_{q,\downarrow}(\bm k^\prime+\bm q)
    ,~\label{eq:Hint_Bloch}
\end{align}
where we define
$
    \tau_{\nu:nm}^{\bm Q}(\bm k,\bm k^\prime) = \bra{u_{n}(\bm k+\bm Q)}\tau_{\nu}\ket{u_{m}(\bm k^\prime)}.
$
In the last equation, we use $\sum_{m}\sum_{\nu}U_\nu\tau_{\nu:nm}^{\bm q}(\bm k,\bm k)[\tau_{\nu:pm}^{\bm q}(\bm k,\bm k)]^*=U\delta_{n,p}$.

To derive the effective Hamiltonian, we evaluate $[\bar{H}_{nmpq}^{\bm Q}(\bm k,\bm k^\prime,\bm q)]_{\bm k_1,\bm k_2}=\bra{\hat{\phi}_{\bm Q:\bm k_1}}\hat{d}_{n,\downarrow}^\dagger(\bm k+\bm q)\hat{d}_{p,\uparrow}^\dagger(\bm k^\prime)\hat{d}_{m,\uparrow}(\bm k)\hat{d}_{q,\downarrow}(\bm k^\prime+\bm q)\ket{\hat{\phi}_{\bm Q:\bm k_2}}$.
Since the states with $n>\tilde{n}$ are unoccupied in $\ket{\hat{\phi}_{\bm Q;\bm k}}$, only the components for $n,m,p,q\leq\tilde{n}$ are nonzero.
Then, it is sufficient to consider the following four cases:

\

{%
\renewcommand*\descriptionlabel[1]{%
  \hspace\labelsep\underline{#1}%
}
\begin{quote}
\begin{description}
  \item[Case (i)]
    $(n<\tilde{n} \lor q<\tilde{n}) \land (m<\tilde{n} \lor p<\tilde{n})$
    \begin{quote}
        Because of $\hat{d}_{n(<\tilde{n}),\sigma}^\dagger(\bm k)\ket{\hat{\phi}_{\bm Q;\bm k^\prime}} = 0$, we obtain
        \begin{align}
            [\bar{H}_{nmpq}^{\bm Q}(\bm k,\bm k^\prime,\bm q)]_{\bm k_1,\bm k_2}
            &=\bra{\hat{\phi}_{\bm Q:\bm k_1}}[\delta_{n,q}\delta_{\bm k,\bm k^\prime}-\hat{d}_{q,\downarrow}(\bm k^\prime+\bm q)\hat{d}_{n,\downarrow}^\dagger(\bm k+\bm q)]
            [\delta_{m,p}\delta_{\bm k,\bm k^\prime}-\hat{d}_{m,\uparrow}(\bm k)\hat{d}_{p,\uparrow}^\dagger(\bm k^\prime)]\ket{\hat{\phi}_{\bm Q:\bm k_2}}
            \notag\\
            &=\delta_{n,q}\delta_{m,p}\delta_{\bm k,\bm k^\prime}\delta_{\bm k_1,\bm k_2}.
        \end{align}
    \end{quote}
        
  \item[Case (ii)]
    $(n<\tilde{n} \lor q<\tilde{n}) \land m=p=\tilde{n} $
    \begin{quote}
        Similarly to the case (i), we obtain
        \begin{align}
            [\bar{H}_{n\tilde{n}\tilde{n}q}^{\bm Q}(\bm k,\bm k^\prime,\bm q)]_{\bm k_1,\bm k_2}
            &=
            \delta_{n,q}\delta_{\bm k,\bm k^\prime}
            [\delta_{\bm k_1,\bm k_2}-\bra{\hat{\phi}_{\bm Q:\bm k_1}}\hat{d}_{\tilde{n},\uparrow}(\bm k)\hat{d}_{\tilde{n},\uparrow}^\dagger(\bm k)\ket{\hat{\phi}_{\bm Q:\bm k_2}}].
        \end{align}
        From the relation
        \begin{align}
            \hat{d}_{\tilde{n},\uparrow}^\dagger(\bm k)\ket{\hat{\phi}_{\bm Q:\bm k_1}} 
            =\hat{d}_{\tilde{n},\uparrow}^\dagger(\bm k)
            \hat{d}_{\tilde{n},\downarrow}^\dagger(\bm k_1+\bm Q)\hat{d}_{\tilde{n},\uparrow}(\bm k_1)
            \ket{\hat{\phi}_{\rm all\uparrow}}
            =-\delta_{\bm k,\bm k_1}\hat{d}_{\tilde{n},\downarrow}^\dagger(\bm k+\bm Q)
            \ket{\hat{\phi}_{\rm all\uparrow}},
        \end{align}
        where we use $\hat{d}_{\tilde{n},\uparrow}^\dagger(\bm k)\ket{\hat{\phi}_{\rm all\uparrow}} = 0$, the second term is given by
        \begin{align}
           \bra{\hat{\phi}_{\bm Q:\bm k_1}}\hat{d}_{\tilde{n},\uparrow}(\bm k)\hat{d}_{\tilde{n},\uparrow}^\dagger(\bm k)\ket{\hat{\phi}_{\bm Q:\bm k_2}} 
           =\delta_{\bm k,\bm k_1}\delta_{\bm k,\bm k_2}\vert\hat{d}_{\tilde{n},\downarrow}^\dagger(\bm k+\bm Q)\ket{\hat{\phi}_{\rm all\uparrow}}\vert^2
           =\delta_{\bm k,\bm k_1}\delta_{\bm k,\bm k_2}.
        \end{align}
        Then, we obtain
        \begin{align}
            [\bar{H}_{n\tilde{n}\tilde{n}q}^{\bm Q}(\bm k,\bm k^\prime,\bm q)]_{\bm k_1,\bm k_2}
            =\delta_{n,q}\delta_{\bm k,\bm k^\prime}\delta_{\bm k_1,\bm k_2}
            [1-\delta_{\bm k,\bm k_1}].
        \end{align}
    \end{quote}
            
  \item[Case (iii)]
    $n = q =\tilde{n} \land (m<\tilde{n} \lor p<\tilde{n})$
    \begin{quote}
        Similarly to the case (i) and (ii), we obtain
        \begin{align}
            [\bar{H}_{\tilde{n}mp\tilde{n}}^{\bm Q}(\bm k,\bm k^\prime,\bm q)]_{\bm k_1,\bm k_2}
            &=
            \delta_{m,p}\delta_{\bm k,\bm k^\prime}
            \bra{\hat{\phi}_{\bm Q:\bm k_1}}\hat{d}_{\tilde{n},\downarrow}^\dagger(\bm k)\hat{d}_{\tilde{n},\downarrow}(\bm k)\ket{\hat{\phi}_{\bm Q:\bm k_2}}.
        \end{align}
        By using the relationship,
        \begin{align}
            \hat{d}_{\tilde{n},\downarrow}(\bm k)\ket{\hat{\phi}_{\bm Q:\bm k_1}}
            =\hat{d}_{\tilde{n},\downarrow}(\bm k)
            \hat{d}_{\tilde{n},\downarrow}^\dagger(\bm k_1+\bm Q)\hat{d}_{\tilde{n},\uparrow}(\bm k_1)
            \ket{\hat{\phi}_{\rm all\uparrow}}
            =\delta_{\bm k,\bm k_1+\bm Q}
            \hat{d}_{\tilde{n},\uparrow}(\bm k-\bm Q)
            \ket{\hat{\phi}_{\rm all\uparrow}},
        \end{align}
        we obtain
        \begin{align}
            [\bar{H}_{\tilde{n}mp\tilde{n}}^{\bm Q}(\bm k,\bm k^\prime,\bm q)]_{\bm k_1,\bm k_2}
            &=\delta_{m,p}\delta_{\bm k,\bm k^\prime}\delta_{\bm k_1,\bm k_2}\delta_{\bm k,\bm k_1+\bm Q}
            \vert\hat{d}_{\tilde{n},\uparrow}(\bm k-\bm Q)
            \ket{\hat{\phi}_{\rm all\uparrow}}
            \vert^2\notag\\
            &=\delta_{m,p}\delta_{\bm k,\bm k^\prime}\delta_{\bm k_1,\bm k_2}\delta_{\bm k,\bm k_1+\bm Q}.
        \end{align}
    \end{quote}
            
  \item[Case (iv)]
    $n = m = p = q =\tilde{n}$
    \begin{quote}
    By using the relationship,
    \begin{align}
        \hat{d}_{\tilde{n},\uparrow}(\bm k)\hat{d}_{\tilde{n},\downarrow}(\bm k^\prime+\bm q)\ket{\hat{\phi}_{\bm Q;\bm k_1}}
        =\delta_{\bm k^\prime+\bm q,\bm k_1+\bm Q}
        \hat{d}_{n,\uparrow}(\bm k)
        \hat{d}_{\tilde{n},\uparrow}(\bm k^\prime+\bm q-\bm Q)\ket{\hat{\psi}_{\rm all\uparrow}},
    \end{align}
    we obtain
    \begin{align}
        &\bra{\hat{\phi}_{\bm Q;\bm k_1}}\hat{d}_{n,\uparrow}^\dagger(\bm k+\bm q)\hat{d}_{\tilde{n},\downarrow}^\dagger(\bm k^\prime)\hat{d}_{\tilde{n},\uparrow}(\bm k)\hat{d}_{\tilde{n},\downarrow}(\bm k^\prime+\bm q)\ket{\hat{\phi}_{\bm Q;\bm k_2}}\notag\\
        &=
        \delta_{\bm k+\bm q,\bm k_1+\bm Q}
        \delta_{\bm k^\prime+\bm q,\bm k_2+\bm Q}
        \bra{\hat{\psi}_{\rm all\uparrow}}
        \hat{d}_{\tilde{n},\uparrow}^\dagger(\bm k+\bm q-\bm Q)
        \hat{d}_{\tilde{n},\uparrow}^\dagger(\bm k^\prime)
        \hat{d}_{\tilde{n},\uparrow}(\bm k)
        \hat{d}_{\tilde{n},\uparrow}(\bm k^\prime+\bm q-\bm Q)
        \ket{\hat{\psi}_{\rm all\uparrow}}\notag\\
        &=
        \delta_{\bm k+\bm q,\bm k_1+\bm Q}
        \delta_{\bm k^\prime+\bm q,\bm k_2+\bm Q}
        (
        \delta_{\bm k,\bm k^\prime}-\delta_{\bm q,\bm Q}
        ).
    \end{align}
    \end{quote}
\end{description}
\end{quote}
}
As a result, we obtain
\begin{align}
  [\mathcal{H}_{\bm Q}^{\rm int}]_{\bm k_1,\bm k_2} &= \sum_\nu\sum_{nmpq}\sum_{\bm k,\bm k^\prime,\bm q}
  \dfrac{U_\nu}{N}\tau_{\nu:nm}^{\bm q}(\bm k,\bm k)
    [\tau_{\nu:qp}^{\bm q}(\bm k^\prime,\bm k^\prime)]^*
    [\bar{H}_{nmpq}^{\bm Q}(\bm k,\bm k^\prime,\bm q)]_{\bm k_1,\bm k_2}\notag\\
    &=\delta_{\bm k_1,\bm k_2}\sum_\nu
    \sum_{n(<\tilde{n}),m(\leq\tilde{n})}\sum_{\bm k,\bm q}
    \dfrac{U_\nu}{N}
    \vert\tau_{\nu:nm}^{\bm q}(\bm k,\bm k)\vert^2
    \notag\\
    &+\delta_{\bm k_1,\bm k_2}\sum_\nu
    \sum_{n(<\tilde{n})}\sum_{\bm q}
    \dfrac{U_\nu}{N}
    \left(
    \vert\tau_{\nu:n\tilde{n}}^{\bm Q}(\bm k_1,\bm q)\vert^2
    -\vert\tau_{\nu:n\tilde{n}}^{\bm q}(\bm k_1,\bm k_1)\vert^2 
    \right)\notag\\
    &+\sum_\nu
    \dfrac{U_\nu}{N}
    \Big(
    \delta_{\bm k_1,\bm k_2}\sum_{n(<\tilde{n})}\sum_{\bm q}
    \vert\tau_{\nu:\tilde{n}\tilde{n}}^{\bm Q}(\bm k_1,\bm q)\vert^2
    -\tau_{\nu:\tilde{n}\tilde{n}}^{\bm Q}(\bm k_1,\bm k_1)[\tau_{\nu:\tilde{n}\tilde{n}}^{\bm Q}(\bm k_2,\bm k_2)]^*
    \Big)\notag\\
    &=
    \sum_\nu
    \dfrac{U_\nu}{N}
    \Big\{
    \delta_{\bm k_1,\bm k_2}\sum_{n(\leq\tilde{n})}\sum_{\bm q}
    \Big(
    \vert\tau_{\nu:n\tilde{n}}^{\bm Q}(\bm k_1,\bm q)\vert^2
    -(1-\delta_{n,\tilde{n}})\vert\tau_{\nu:n\tilde{n}}^{\bm q}(\bm k_1,\bm k_1)\vert^2 
    \Big)
    -\tau_{\nu:\tilde{n}\tilde{n}}^{\bm Q}(\bm k_1,\bm k_1)[\tau_{\nu:\tilde{n}\tilde{n}}^{\bm Q}(\bm k_2,\bm k_2)]^*
    \Big\}\notag\\
    &+\bra{\hat{\phi}_{\rm all\uparrow}}\hat{H}_{\rm int}\ket{\hat{\phi}_{\rm all\downarrow}},~\label{eq:Heffint_Bloch}
\end{align}
where we have used $\bra{\hat{\phi}_{\rm all\uparrow}}\hat{H}_{\rm int}\ket{\hat{\phi}_{\rm all\downarrow}}=\sum_{n(<\tilde{n}),m(\leq\tilde{n})}\sum_{\bm k,\bm q}\frac{U_\nu}{N}\vert\tau_{\nu:nm}^{\bm q}(\bm k,\bm k)\vert^2$.
When the isolated band is the lowest band,
by using $D_{nmpq}^{\nu:\bm Q}(\bm k_1,\bm k_2,\bm k_3,\bm k_4)
=1-\tau_{\nu:nm}^{\bm Q}(\bm k_1,\bm k_2)[\tau_{\nu:pq}^{\bm Q}(\bm k_4,\bm k_3)]^*$ and $\bra{\hat{\phi}_{\rm all\uparrow}}\hat{H}_{\rm int}\ket{\hat{\phi}_{\rm all\downarrow}}=0$,
we obtain
\begin{align}
    [\mathcal{H}_{\bm Q}^{\rm int}]_{\bm k_1,\bm k_2}=\sum_{\nu}
    \dfrac{U_\nu}{N}
    \left(
    \sum_{\bm q}
    \delta_{\bm k_1,\bm k_2}
    [1-D_{{\tilde{n}}{\tilde{n}}{\tilde{n}}{\tilde{n}}}^{\nu:\bm Q}(\bm k_1,\bm q,\bm q,\bm k_1)]
    -[1-D_{{\tilde{n}}{\tilde{n}}{\tilde{n}}{\tilde{n}}}^{\nu:\bm Q}(\bm k_1,\bm k_1,\bm k_2,\bm k_2)]
    \right),
\end{align}
which is the formula shown in the main text.

\subsubsection{Gauge invariance}

The Bloch wave function has gauge degrees of freedom.
Under the gauge transformation, $\ket{u_n(\bm k)}\rightarrow\ket{u_n(\bm k)}e^{i\kappa_n(\bm k)}$, $\mathcal{H}_{\bm Q}^{\rm int}$ is transformed as
\begin{align}
    [\mathcal{H}_{\bm Q}^{\rm int}]_{\bm k_1,\bm k_2}\rightarrow
    [\mathcal{H}_{\bm Q}^{\rm int}]_{\bm k_1,\bm k_2}
    e^{-i(\kappa_{\tilde{n}}(\bm k_1+\bm Q)-\kappa_{\tilde{n}}(\bm k_1))}e^{i(\kappa_{\tilde{n}}(\bm k_2+\bm Q)-\kappa_{\tilde{n}}(\bm k_2))}.
\end{align}
Therefore, $\mathcal{H}_{\bm Q}^{\rm int}$ is clearly gauge covariant.
However, this transformation is rewritten as
\begin{align}
    \mathcal{H}_{\bm Q}^{\rm int}
    \rightarrow U_\kappa\mathcal{H}_{\bm Q}^{\rm int}U_\kappa^\dagger,
\end{align}
with $U_\kappa = {\rm diag}(\cdots,e^{-i(\kappa_n(\bm k+\bm Q)-\kappa_n(\bm k))}\cdots)$ which is the unitary matrix.
Thus, the gauge transformation of $\mathcal{H}_{\bm Q}^{\rm int}$ is equivalent to the unitary transformation.
In addition, because of $[\mathcal{H}_{\bm Q}^{0} , U_\kappa] = 0$, the gauge transformation reads
\begin{align}
 \mathcal{H}_{\bm Q}
    \rightarrow U_\kappa\mathcal{H}_{\bm Q}U_\kappa^\dagger,
\end{align}
which is also regarded as a unitary transformation.
Thus, the eigenvalue of $\mathcal{H}_{\bm Q}$ is gauge independent.

\subsection{Spin stifness}

The spin stiffness is defined by $\mathcal{D}_{\rm spin}^{ab} = \lim_{\bm Q \to0}\partial_{Q_a} \partial_{Q_b}E_{\rm spin}(\bm Q)$, where $a$ and $b$ denote the spatial directions.
Since $\mathcal{D}_{\rm spin}^{ab}$ is a symmetric matrix, there should be a coordinate system such that $\mathcal{D}_{\rm spin}^{ab} = 0$ for $a \neq b$. Therefore, we focus on the diagonal component of the spin stiffness, i.e., $\mathcal{D}_{\rm spin}^{aa}$.
By using $\bm A_{\bm 0}^0 = (1,1,\cdots,1)^{\mathsf{T}}/\sqrt{N}$ and $[\mathcal{H}_{\bm Q}-E_{\rm FM}]\bm A^0 =0$ due to 
$SU(2)$ symmetry,
$\mathcal{D}_{\rm spin}^{aa}$ is given by
\begin{align}
    \mathcal{D}_{\rm spin}^{aa} 
    &=\lim_{\bm Q \to0}\partial_{Q_a}^2\left(
    [\bm A_{\bm 0}^0]^\dagger
    [\mathcal{H}_{\bm Q}-E _{\rm FM}]\bm A_{\bm 0}^0
    \right)
    \notag\\
    &= \lim_{\bm Q \to0}
    \left([\bm A_{\bm 0}^0]^\dagger\partial_{Q_a} ^2
    \mathcal{H}_{\bm Q}\bm A_{\bm 0}^0
    + 2\partial_{Q_a}[\bm A_{\bm 0}^0]^\dagger 
    [\mathcal{H}_{\bm Q}-E_{\rm FM}]
    \partial_{Q_a}\bm A_{\bm Q}^0
    +2\partial_{Q_a}[\bm A_{\bm 0}^0]^\dagger 
    \partial_{Q_a}\mathcal{H}_{\bm Q}
    \bm A_{\bm 0}^0
    +2[\bm A_{\bm 0}^0]^\dagger 
    \partial_{Q_a}\mathcal{H}_{\bm Q}
    \partial_{Q_a}\bm A_{\bm 0}^0
    \right)\notag\\
    &= \lim_{\bm Q \to0}
    \left[\dfrac{1}{N}\sum_{\bm k_1,\bm k_2}[\partial_{Q_a} ^2
    \mathcal{H}_{\bm Q}]_{\bm k_1,\bm k_2}
    - \sum_{\chi\neq0}\dfrac{2}{E^\chi_{\rm spin}(\bm 0)}\Big\vert[\bm A_{\bm 0}^0]^\dagger 
    \partial_{Q_a}\mathcal{H}_{\bm Q}
    \bm A_{\bm 0}^{\chi}\Big\vert^2
    \right].~\label{eq:Dspin}
\end{align}
To obtain the second equality, we have used the Hellmann-Feynman theorem that $\bm A_{\bm Q}^\chi$ satisfies.
In the following, we rewrite this formula by using quantum geometry and energy dispersion.

\subsubsection{Expansion of the effective Hamiltonian}

To derive the formula in terms of the quantum geometry and energy dispersion, we expand $\mathcal{H}_{\bm Q}^0$ and $\mathcal{H}_{\bm Q}^{\rm int}$, up to $O(Q^2)$.
The noninteracting term is easily expanded as
\begin{align}
    [\mathcal{H}_{\bm Q}^{0}]_{\bm k,\bm k^\prime} = \delta_{\bm k,\bm k^\prime}[E_{\rm FM} + Q_a\partial_{k_a}\epsilon_{\tilde{n}}(\bm k)+ Q_aQ_b\partial_{k_a}\partial_{k_b}\epsilon_{\tilde{n}}(\bm k)/2].~\label{eq:Heff_ene_velo}
\end{align}
In contrast, the calculation of the Coulomb interaction term is more involved. We first expand
$
    \tau_{\nu:nm}^{\bm Q}(\bm k,\bm k^\prime) = \bra{u_{n}(\bm k+\bm Q)}\tau_{\nu}\ket{u_{m}(\bm k^\prime)}.
$
It is helpful to introduce the following gauge fixed quantity~\cite{SDaido2024}:
\begin{align}
    \tilde{\tau}_{\nu:nm}^{\bm Q}(\bm k,\bm k^\prime) &= e^{i\theta_n(\bm k;\bm Q)}\tau_{\nu:nm}^{\bm Q}(\bm k,\bm k^\prime),\\
    \theta_{n}(\bm k;\bm Q) &=-i\int_{0}^{\bm Q}d\bar{\bm Q}\cdot\braket{u_n(\bm k+\bar{\bm Q})\vert \bm \partial_{\bm Q}u_n(\bm k+\bar{\bm Q})},
\end{align}
with Wilson line $\theta_n(\bm k;\bm Q)$ where the integral is taken along the straight line $\bm 0 \rightarrow \bm Q$.
Up to $O(Q^2)$, the expansion of $\tilde{\tau}_{\nu:nm}^{\bm Q}(\bm k,\bm k^\prime)$ is transparent from the viewpoint of the gauge covariance: 
\begin{align}
    &\tilde{\tau}_{\nu:{np}}^{\bm Q}(\bm k,\bm k^\prime) 
    =\underbrace{\Bigg\{1 
    + i\bigg[1+\dfrac{\bm Q\cdot\bm \partial_{\bm k}}{2}\bigg]\bm A_{nn}(\bm k)\cdot \bm Q 
    + \dfrac{[i\bm A_{nn}(\bm k)\cdot \bm Q]^2}{2}
    \Bigg\}}_{{\rm Expansion\ of}\ e^{i\theta_n(\bm k;\bm Q)}}\notag\\
    &\times
    \underbrace{\sum_{p}\Bigg\{\delta_{n,p} - i\bigg[1+\dfrac{\bm Q\cdot\bm \partial_{\bm k}}{2}\bigg]\bm A_{{n}p}(\bm k)\cdot\bm Q 
    +\sum_{q}\dfrac{[i\bm A_{{n}q}(\bm k)\cdot \bm Q]
    [i\bm A_{qp}(\bm k)\cdot \bm Q]}{2}
    \Bigg\}\tau_{\nu:pm}(\bm k,\bm k^\prime)}_{{\rm Expansion\ of}\ \tau_{\nu:nm}^{\bm Q}(\bm k,\bm k^\prime)}+O(Q^3)\notag\\
    &=\sum_{p}\Bigg[
    \delta_{{n},p}-(1-\delta_{n,p})i\bm A_{{n}p}(\bm k)\cdot\bm Q
    -\sum_{q(\neq {n},p)}
    \dfrac{[\bm A_{{n}q}(\bm k)\cdot \bm Q]
    [\bm A_{qp}(\bm k)\cdot \bm Q]}{2}
    \notag\\
    &-\dfrac{i}{2}(1-\delta_{{n},p})
    \Bigg\{
    [\bm Q\cdot\bm \partial_{\bm k}]
    [\bm A_{{n}p}(\bm k)\cdot\bm Q ]
    +i\big[\bm A_{nn}(\bm k)\cdot \bm Q-\bm A_{pp}(\bm k)\cdot \bm Q\big]
    [\bm A_{{n}p}(\bm k)\cdot \bm Q]
    \Bigg\}
    \Bigg]
    \tau_{\nu:pm}(\bm k,\bm k^\prime)+O(Q^3),
\end{align}
with $\tau_{\nu:nm}(\bm k,\bm k^\prime) \equiv \tau_{\nu:nm}^{\bm 0}(\bm k,\bm k^\prime)$ and the Berry connection $A_{nm}^{a}(\bm k) = -i\braket{u_n(\bm k)\vert \partial_{k_a} u_m(\bm k)}$.
This is rewritten in terms of the quantum geometric quantities as
\begin{align}
   \tilde{\tau}_{\nu:{nm}}^{\bm Q}(\bm k,\bm k^\prime) =
   \sum_{p} \left\{\delta_{{n},p}\left[1-\dfrac{g_{n}^{ab}(\bm k)}{2}Q_aQ_b\right]
   +(1-\delta_{{n},p})
   \left[\dfrac{R_{{n}p}^{ab}(\bm k)-f_{{n}p}^{ab}(\bm k)}{2}Q_aQ_b
   - iA_{{n}m}^a(\bm k)Q_a
   \right]
   \right\}
   \tau_{\nu:pm}(\bm k,\bm k^\prime)\notag\\
   +O(Q^3),
\end{align}
with the following quantum geometric quantities:
\begin{align}
    g_{n}^{ab}(\bm k) &= \sum_{m(\neq n)}\dfrac{A_{nm}^{a}(\bm k)A_{mn}^{b}(\bm k)+A_{nm}^{b}(\bm k)A_{mn}^{a}(\bm k)}{2},\\
    f_{nm}^{ab}(\bm k) &= \sum_{p(\neq n,m)}\dfrac{A_{np}^{a}(\bm k)A_{pm}^{b}(\bm k)+A_{np}^{b}(\bm k)A_{pm}^{a}(\bm k)}{2},\\
    R_{nm}^{ab}(\bm k) &=-i\mathscr{D}_a[A_{nm}^b(\bm k)]
    \notag\\
    &=-i[\partial_{k_a}A_{nm}^b(\bm k)+i(A_{nn}^a(\bm k)-A_{mm}^a(\bm k))A_{nm}^b(\bm k)],
\end{align}
where $\mathscr{D}_a[\mathcal{O}]$ is the gauge covariant derivative.
In particular, $g_{n}^{ab}(\bm k)$ is known as the quantum metric and $R_{nm}^{ab}(\bm k)$ is closely related to the shift vector~\cite{SSipe2000}.
$f_{nm}^{ab}(\bm k)$ is the quantity resembling a quantum metric.
As a result, the quantum geometric formula of $\tau_{\nu:nm}^{\bm Q}(\bm k,\bm k^\prime)$ is defined by $\tau_{\nu:nm}^{\bm Q:{\rm g}}(\bm k,\bm k^\prime)$ as follows:
\begin{align}
    \tau_{\nu:nm}^{\bm Q}(\bm k,\bm k^\prime) = e^{-i\theta_{n}(\bm k;\bm Q)}[\tau_{\nu:nm}(\bm k,\bm k^\prime) + \tau_{\nu:nm}^{\bm Q:{\rm g}}(\bm k,\bm k^\prime)].~\label{eq:tau_qg}
\end{align}
By inserting Eq.~\eqref{eq:tau_qg} into Eq.~\eqref{eq:Heffint_Bloch}, the Coulomb interaction term of the effective Hamiltonian is given by
\begin{align}
    \mathcal{H}_{\bm Q}^{\rm int}&=U_{\theta:\bm Q}\bar{\mathcal{H}}_{\bm Q}^{\rm geom}U_{\theta:\bm Q}^\dagger
    +O(Q^3)\label{eq:Heff_int_geom}\\
    [\bar{\mathcal{H}}_{\bm Q}^{\rm geom}]_{\bm k_1,\bm k_2}
    &=\sum_{\nu}\dfrac{U_\nu}{N}
    \Bigg\{
    \delta_{\bm k_1,\bm k_2}\sum_{n(\leq\tilde{n})}\sum_{\bm q}
    \bigg[
    \vert\tau_{\nu:n\tilde{n}}(\bm k_1,\bm q)\vert^2 
    -(1-\delta_{n,\tilde{n}})\vert\tau_{\nu:\tilde{n}n}(\bm k_1,\bm q)\vert^2 
    \notag\\
    &
    + \left(\tau_{\nu:n\tilde{n}}^{\bm Q:{\rm g}}(\bm k_1,\bm q)[\tau_{\nu:n\tilde{n}}(\bm k_1,\bm q)]^*+c.c.\right)
    +\vert\tau_{\nu:n\tilde{n}}^{\bm Q:{\rm g}}(\bm k_1,\bm k)\vert^2
    \bigg]
    -\tau_{\nu:\tilde{n}\tilde{n}}(\bm k_1,\bm k_1)\tau_{\nu:\tilde{n}\tilde{n}}(\bm k_2,\bm k_2) 
    \notag\\
    &
    -\left(\tau_{\nu:\tilde{n}\tilde{n}}^{\bm Q:{\rm g}}(\bm k_1,\bm k_1)[\tau_{\nu:\tilde{n}\tilde{n}}(\bm k_2,\bm k_2)]^*
    +\tau_{\nu:\tilde{n}\tilde{n}}(\bm k_1,\bm k_1)[\tau_{\nu:\tilde{n}\tilde{n}}^{\bm Q:{\rm g}}(\bm k_2,\bm k_2)]^*
    \right)
    -\tau_{\nu:\tilde{n}\tilde{n}}^{\bm Q:{\rm g}}(\bm k_1,\bm k_1)[\tau_{\nu:\tilde{n}\tilde{n}}^{\bm Q:{\rm g}}(\bm k_2,\bm k_2)]^*
    \Bigg\},
\end{align}
with $U_{\theta:\bm Q} = {\rm diag}(\cdots,e^{-i\theta_{\tilde{n}}(\bm k;\bm Q)},\cdots)$.
By summing up Eqs.~\eqref{eq:Heff_int_geom} and~\eqref{eq:Heff_ene_velo}, we obtain
\begin{align}
    \mathcal{H}_{\bm Q} &= \mathcal{H}_{\bm Q}^{0} + U_{\theta:\bm Q}\bar{\mathcal{H}}_{\bm Q}^{\rm geom}U_{\theta:\bm Q}^\dagger + O(Q^3)\notag\\
    &=U_{\theta:\bm Q}(\mathcal{H}_{\bm Q}^{0} + \bar{\mathcal{H}}_{\bm Q}^{\rm geom})U_{\theta:\bm Q}^\dagger + O(Q^3)\notag\\
    &=U_{\theta:\bm Q}\bar{\mathcal{H}}_{\bm Q}U_{\theta:\bm Q}^\dagger + O(Q^3),
\end{align}
with $\bar{\mathcal{H}}_{\bm Q} = \mathcal{H}_{\bm Q}^{0} + \bar{\mathcal{H}}_{\bm Q}^{\rm geom}$.
In the second equality, we have used $[\mathcal{H}_{\bm Q}^{0}, U_{\theta:\bm Q}]=0$, and $U_{\theta:\bm Q}^\dagger U_{\theta:\bm Q}=\bm 1$.

\subsubsection{Gauge invariant formula}
Then, we derive an explicitly gauge-invariant formula for the spin stiffness.
The first term of Eq.~\eqref{eq:Dspin} is rewritten as
\begin{align}
    \lim_{\bm Q \to 0}
    [\bm A_{\bm 0}^0]^\dagger\partial_{Q_a} ^2
    \mathcal{H}_{\bm Q}\bm A_{\bm 0}^0
    &=\lim_{\bm Q \to 0}
    [\bm A_{\bm 0}^0]^\dagger\partial_{Q_a} ^2
    \left(\mathcal{H}_{\bm Q}-E_{\rm FM}U_{\theta:\bm Q}U_{\theta:\bm Q}^\dagger\right)\bm A_{\bm 0}^0\notag\\
    &=\lim_{\bm Q \to 0}[\bm A_{\bm 0}^0]^\dagger\left(
    \partial_{Q_a} ^2
    \bar{\mathcal{H}}_{\bm Q}
    +2(\partial_{Q_a}U_{\theta:\bm Q} \partial_{Q_a}
    \bar{\mathcal{H}}_{\bm Q}+ {\rm c.c.})
    +2\partial_{Q_a}U_{\theta:\bm Q}
    [\mathcal{H}_{\bm 0}-E_{\rm FM}]
    \partial_{Q_a}U_{\theta:\bm Q}^\dagger
    \right)\bm A_{\bm 0}^0,~\label{eq:Dspin_first}
\end{align}
where we use $U_{\theta:\bm Q}\xrightarrow{\bm Q\to0}\bm 1$ and $[\bar{\mathcal{H}}_{\bm 0}-E_{\rm FM}]\bm A_{\bm 0}^0=0$ because of $\bar{\mathcal{H}}_{\bm 0}=\mathcal{H}_{\bm 0}$.
The second term of Eq.~\eqref{eq:Dspin} is rewritten by
\begin{align}
    -\lim_{\bm Q\to0}\sum_{\chi\neq0}
    &\dfrac{2}{E^\chi_{\rm spin}(\bm 0)}
    \Big\vert[\bm A_{\bm 0}^0]^\dagger 
    \partial_{Q_a}\mathcal{H}_{\bm Q}
    \bm A_{\bm 0}^{\chi}\Big\vert^2
    =-\lim_{\bm Q\to0}\sum_{\chi\neq0}
    \dfrac{2}{E^\chi_{\rm spin}(\bm 0)}
    \Big\vert[\bm A_{\bm 0}^0]^\dagger 
    \partial_{Q_a}[\mathcal{H}_{\bm Q}-E_{\rm FM}U_{\theta:\bm Q}U_{\theta:\bm Q}^\dagger]
    \bm A_{\bm 0}^{\chi}\Big\vert^2
    \notag\\
    =&-\lim_{\bm Q \to0}\sum_{\chi\neq0}\dfrac{2}{E^\chi_{\rm spin}(\bm 0)}\Big\vert[
    \bm A_{\bm 0}^0]^\dagger
    \partial_{Q_a}U_{\theta:\bm Q}\underbrace{[\bar{\mathcal{H}}_{\bm 0}-E_{\rm FM}] 
    \bm A_{\bm 0}^{\chi}}_{[\mathcal{H}_{\bm 0}-E_{\rm FM}] 
    \bm A_{\bm 0}^{\chi}=E_{\rm spin}^\chi(\bm 0) 
    \bm A_{\bm 0}^{\chi}}
    +[\bm A_{\bm 0}^0]^\dagger\partial_{Q_a}\bar{\mathcal{H}}_{\bm Q}
    \bm A_{\bm 0}^{\chi}
    +
    \underbrace{
    [\bm A_{\bm 0}^0]^\dagger[\bar{\mathcal{H}}_{\bm 0}-E_{\rm FM}] 
    }_{[\bm A_{\bm 0}^0]^\dagger[\bar{\mathcal{H}}_{\bm 0}-E_{\rm FM}]
    =0}    
    \partial_{Q_a}U_{\theta:\bm Q}^\dagger
    \bm A_{\bm 0}^{\chi}
    \Big\vert^2
    \notag\\
    =&-2\lim_{\bm Q \to0}
    \sum_{\chi\neq0}\dfrac{1}{E^\chi_{\rm spin}(\bm 0)}
    \Big\vert[\bm A_{\bm 0}^0]^\dagger \partial_{Q_a}\bar{\mathcal{H}}_{\bm Q}
    \bm A_{\bm 0}^{\chi}\Big\vert^2
    \notag\\
    &    
    -2\lim_{\bm Q \to0}
    \bigg\{
    \big(\bm [\bm A_{\bm 0}^0]^\dagger \partial_{Q_a}\bar{\mathcal{H}}_{\bm Q}
    \underbrace{\sum_{\chi}
    \bm A_{\bm 0}^{\chi}
    [\bm A_{\bm 0}^\chi]^\dagger
    }_{\bm 1}
    \partial_{Q_a}U_{\theta:\bm Q}^\dagger\bm A_{\bm 0}^0+ {\rm c.c.}\big)
    +[\bm A_{\bm 0}^0]^\dagger\partial_{Q_a}U_{\theta:\bm Q}
    \underbrace{\sum_{\chi}E^\chi_{\rm spin}(\bm 0)
    \bm A_{\bm 0}^{\chi}
    [\bm A_{\bm 0}^\chi]^\dagger
    }_{\bar{\mathcal{H}}_{\bm 0}-E_{\rm FM}}
    \partial_{Q_a}U_{\theta:\bm Q}^\dagger\bm A_{\bm 0}^0
    \bigg\}\notag\\
    =&-2\lim_{\bm Q \to0}\left(\sum_{\chi\neq0}\dfrac{\vert\bm A^\dagger\partial_{Q_a}\bar{\mathcal{H}}_{\bm Q}
    \bm A_{\bm 0}^{\chi}\vert^2}{E^\chi_{\rm spin}(\bm 0)}
    +([\bm A_{\bm 0}^0]^\dagger\partial_{Q_a}\bar{\mathcal{H}}_{\bm Q}
    \partial_{Q_a}U_{\theta:\bm Q}^\dagger\bm A_{\bm 0}^0+{\rm c.c.})
    +[\bm A_{\bm 0}^0]^\dagger\partial_{Q_a}U_{\theta:\bm Q}[\mathcal{H}_{\bm 0}-E_{\rm FM}]\partial_{Q_a}U_{\theta:\bm Q}^\dagger\bm A_{\bm 0}^0
    \right).~\label{eq:Dspin_second}
\end{align}
In the second equality, we have used $\bm [A_{\bm 0}^0]^\dagger\partial_{Q_a}\bar{\mathcal{H}}_{\bm Q}A_{\bm 0}^0[A_{\bm 0}^0]^\dagger\partial_{Q_a}U_{\theta:\bm Q}^\dagger A_{\bm 0}^0+{\rm c.c.}\xrightarrow{\bm Q\to0}0$ and $E_{\rm spin}^0(\bm 0) = 0$.
By summing up Eqs.~\eqref{eq:Dspin_first} and ~\eqref{eq:Dspin_second}, we obtain the explicitly gauge invariant formula as,
\begin{align}
    D_{\rm spin}^{aa} = 
    \lim_{\bm Q \to0}
    \left(
    [\bm A_{\bm 0}^0]^\dagger
    \partial_{Q_a} ^2
    \bar{\mathcal{H}}_{\bm Q}
    \bm A_{\bm 0}^0
    -2\sum_{\chi\neq0}\dfrac{\vert\bm [\bm A_{\bm 0}^0]^\dagger\partial_{Q_a}\bar{\mathcal{H}}_{\bm Q}
    \bm A_{\bm 0}^{\chi}\vert^2}{E^\chi_{\rm spin}(\bm 0)}
    \right).~\label{eq:Dspin_gauge_invaariant}
\end{align}

\subsubsection{Quantum geometry and energy dispersion}
Finally, we derive the formula for the spin stiffness in terms of the quantum geometry and the energy dispersion.
The first term of Eq.~\eqref{eq:Dspin_gauge_invaariant}, which arises from the acoustic magnon, is given by,
\begin{align}
    \mathcal{D}_{\rm acou}^{aa} &= \lim_{\bm Q \to0}[\bm A_{\bm 0}^0]^\dagger\partial_{Q_a} ^2
    \bar{\mathcal{H}}_{\bm Q}\bm A_{\bm 0}^0\notag\\
    &=\lim_{\bm Q \to0}\dfrac{1}{N}\sum_{\bm k,\bm k^\prime}\partial_{Q_a} ^2[\bar{\mathcal{H}}_{\bm Q}]_{\bm k,\bm k^\prime}.
\end{align}
This is divided into effective-mass and quantum geometric contributions as,
\begin{align}
    \mathcal{D}_{\rm acou}^{aa} &= \mathcal{D}_{\rm mass}^{aa} + \mathcal{D}_{\rm geom}^{aa},\\
    \mathcal{D}_{\rm mass}^{aa} &= \lim_{\bm Q \to0}\dfrac{1}{N}\sum_{\bm k,\bm k^\prime}\partial_{Q_a} ^2[\mathcal{H}_{\bm Q}^{0}]_{\bm k,\bm k^\prime},\\
    \mathcal{D}_{\rm geom}^{aa} &= \lim_{\bm Q \to0}\dfrac{1}{N}\sum_{\bm k,\bm k^\prime}\partial_{Q_a} ^2[\bar{\mathcal{H}}_{\bm Q}^{\rm geom}]_{\bm k,\bm k^\prime}.
\end{align}
Because of the periodicity, we easily find that
\begin{align}
    \mathcal{D}_{\rm mass}^{aa}
    =\dfrac{1}{N}\sum_{\bm k}\partial_{k_a} ^2\epsilon_{\tilde{n}}(\bm k) = 0.
\end{align}
Thus, the acoustic magnon contribution is given only by the quantum geometric contribution:
\begin{align}
    \mathcal{D}_{\rm geom}^{aa}
    &=\sum_{\nu}\dfrac{U_\nu}{N^2}\sum_{\bm k}\tau_{\nu:{\tilde{n}\tilde{n}}}(\bm k)\sum_{\bm k}\left\{g_{\tilde{n}}^{aa}(\bm k)\tau_{\nu:{\tilde{n}\tilde{n}}}(\bm k)
    +\sum_{m(\neq{\tilde{n}})}(f_{{\tilde{n}}m}^{aa}(\bm k)-R_{{\tilde{n}}m}^{aa}(\bm k))\tau_{\nu:m{\tilde{n}}}(\bm k)+c.c.
    \right\}\notag\\
    &-\sum_{\nu}\dfrac{U_\nu}{N^2}\vert\sum_{\bm k}\sum_{n(\neq{\tilde{n}})}A_{{\tilde{n}}n}^a(\bm k)\tau_{\nu:n{\tilde{n}}}(\bm k)\vert^2
    + \lim_{\rm \bm Q\rightarrow0}\partial_{Q_a}^2\sum_{\nu}\dfrac{U_\nu}{N^2}\sum_{n(\leq\tilde{n})}\sum_{\bm k,\bm q}\vert\tau_{\nu:n\tilde{n}}^{\bm Q}(\bm k,\bm q)\vert^2\notag\\
    &=\sum_{\nu}\dfrac{U_\nu}{N^2}\sum_{\bm k}\tau_{\nu:{\tilde{n}\tilde{n}}}(\bm k)\sum_{\bm k}\left\{g_{\tilde{n}}^{aa}(\bm k)\tau_{\nu:{\tilde{n}\tilde{n}}}(\bm k)
    +\sum_{m(\neq{\tilde{n}})}(f_{{\tilde{n}}m}^{aa}(\bm k)-R_{{\tilde{n}}m}^{aa}(\bm k))\tau_{\nu:m{\tilde{n}}}(\bm k)+c.c.
    \right\}\notag\\
    &-\sum_{\nu}\dfrac{U_\nu}{N^2}\vert\sum_{\bm k}\sum_{n(\neq{\tilde{n}})}A_{{\tilde{n}}n}^a(\bm k)\tau_{\nu:n{\tilde{n}}}(\bm k)\vert^2
    + \sum_{\nu}\dfrac{U_\nu}{N^2}\sum_{n(\leq\tilde{n})}\sum_{\bm k,\bm q}\partial_{k_a}^2\vert\tau_{\nu:n\tilde{n}}(\bm k,\bm q)\vert^2\notag\\
    &= \sum_{\nu}\dfrac{U_\nu}{N^2}\sum_{\bm k}\tau_{\nu:{\tilde{n}\tilde{n}}}(\bm k)\sum_{\bm k}\left\{g_{\tilde{n}}^{aa}(\bm k)\tau_{\nu:{\tilde{n}\tilde{n}}}(\bm k)
    +\sum_{m(\neq{\tilde{n}})}(f_{{\tilde{n}}m}^{aa}(\bm k)-R_{{\tilde{n}}m}^{aa}(\bm k))\tau_{\nu:m{\tilde{n}}}(\bm k)+c.c.
    \right\}\notag\\
    &-2\sum_{\nu}\dfrac{U_\nu}{N^2}\vert\sum_{\bm k}\sum_{n(\neq{\tilde{n}})}A_{{\tilde{n}}n}^a(\bm k)\tau_{\nu:n{\tilde{n}}}(\bm k)\vert^2,
\end{align}
with $\tau_{\nu:nm}(\bm k) = \tau_{\nu:nm}(\bm k,\bm k)$.

The second term of Eq~\eqref{eq:Dspin_gauge_invaariant} arises from transitions between the acoustic-magnon state and the high-energy excited states, such as states corresponding to the Stoner continuum and the optical magnon,
\begin{align}
    \mathcal{D}_{\rm hexc}^{aa}=-2\lim_{\bm Q\to0}\sum_{\chi\neq0}\dfrac{\vert\bm [\bm A_{\bm 0}^0]^\dagger\partial_{Q_a}\bar{\mathcal{H}}_{\bm Q}
    \bm A_{\bm 0}^{\chi}\vert^2}{E^\chi_{\rm spin}(\bm 0)}.
\end{align}
The transition from the acoustic-magnon state to the high-energy states is mediated by the first derivative of $\bar{\mathcal{H}}_{\bm Q}$,
\begin{align}
    [\lim_{\bm Q \rightarrow0}\partial_{Q_a}\bar{\mathcal{H}}_{\bm Q}]_{\bm k,\bm k^\prime}
    &= \delta_{\bm k,\bm k^\prime}\partial_{k_a}[\epsilon_{\tilde{n}}(\bm k)
    +\sum_{\nu}\dfrac{U_\nu}{N}\sum_{n(\leq\tilde{n})}\sum_{\bm q}\vert\tau_{\nu:{n\tilde{n}}}(\bm k,\bm q)\vert^2]\notag\\
    &+i\sum_{\nu}\dfrac{U_\nu}{N}\sum_{n(\neq{\tilde{n}})}
    \left[A^{a}_{{\tilde{n}}n}(\bm k)\tau_{\nu:n{\tilde{n}}}(\bm k)\tau_{\nu:{\tilde{n}\tilde{n}}}(\bm k^\prime)
    -\tau_{\nu:{\tilde{n}}\tilde{n}}(\bm k)\tau_{\nu:{\tilde{n}}n}(\bm k^\prime)[A^{a}_{n{\tilde{n}}}(\bm k^\prime)]
    \right].~\label{eq:Hq_1deriv}
\end{align}
Therefore, in contrast to the quantum geometric contribution,  $\mathcal{D}_{\rm hexc}^{aa}$ includes the energy dispersion contribution via its group velocity, $\partial_{k_a}\epsilon_{\tilde{n}}(\bm k)$.
As a result, we obtain the formula for the spin stiffness in terms of the quantum geometry and the energy dispersion as 
\begin{align}
    \mathcal{D}^{aa}_{\rm spin} =\mathcal{D}^{aa}_{\rm geom} +\mathcal{D}^{aa}_{\rm hexc},
\end{align}
with Eq.~\eqref{eq:Hq_1deriv}.

\section{Hubbard models with tunable quantum geometry on extended Tasakai's lattice}

We construct and study the Hubbard model with a saturated ferromagnetic ground state, in which the quantum geometry is tuned independently of the energy dispersion and Coulomb interaction.
While we study quasi one-dimensional systems in the main text, here, we consider arbitrary dimensions, including one, two, and three dimensions.

\subsection{$d$-dimensional extended Tasaki lattice}

We begin by defining the extended Tasaki lattice on which we construct the Hubbard model.
Following Refs.~\cite{STasaki1995,STasaki2020}, we first introduce the original $d$-dimensional Tasaki lattice. 
The Tasaki lattice consists of a $d$-dimensional Hypercubic lattice with additional sites located at the midpoints of the bonds between nearest-neighbor sites. 
While the original Tasaki lattice can be defined in any dimension, here we focus on $d\leq 3$, for simplicity. Note that its extension to higher dimensions is straightforward.

The $d$-dimensional hypercubic lattice is defined in the Cartesian coordinate system, with the set of sites given by
\begin{align}
    \mathscr{E}_{d}\equiv \Big\{\bm R=(R_1,\cdots,R_d)\big\vert R_a\in\{0,1,2,\cdots, L-1\}\Big\},
\end{align}
where $L$ is a positive integer. 
We impose periodic boundary conditions, and the total number of sites is $N=L^d$.
Additional sites are placed at the midpoints of the edges of the hypercubic lattice. More precisely, we define the set of these sites by 
\begin{align}
    \mathscr{I}_{d}
    \equiv
    \bigcup_{a=1}^{d}
    \Big\{
        \bm R + \frac{1}{2}\bm e_a
        \,\Big|\,
        \bm R \in \mathscr{E}_{d}
    \Big\},
\end{align}
where $\bm e_a$ denotes the unit vector in the $a$th Cartesian direction.

While the original Tasaki lattice is defined  by $\mathscr{E}_{d} \cup \mathscr{I}_{d}$, we further introduce the following $(N_{\rm sub} -1)$ sets of sites
\begin{align}
    \mathscr{I}_{d}^i
    \equiv
    \Big\{
        \bm R + \tilde{\bm\mu}_i
        \,\Big|\,
        \bm R \in \mathscr{I}_{d}
    \Big\},
\end{align}
with $i = 1,\ldots,N_{\rm sub}-1$ where $\tilde{\bm\mu}_i$ are arbitrary $d$-dimensional vectors. 
Then, we define the extended Tasaki Lattice by
\begin{align}
    \Lambda_d \equiv \mathscr{E}_d\cup\mathscr{I}_{d}^1\cup\cdots\cup\mathscr{I}_{d}^{N_{\rm sub}-1}.
\end{align}
The number of total sites and the number of sublattices are given by $N + Nd(N_{\rm sub}-1)$ and $1 + d(N_{\rm sub}-1)$, respectively.
In particular, for $d=1$, by assuming $[\tilde{\bm \mu}_i]_1= 0$, $\Lambda_d$ corresponds to the set of 1D periodic chains in the main text.

\subsection{Hubbard model with tunable quantum geometry~\label{sec:Hubbar_qg}}

We define the fermionic annihilation operator of spin $\sigma$ at the sites in $\mathscr{E}_{d}$ and $\mathscr{I}_{d}^i$ as
$\hat{c}_{\bm x,\sigma
}$ and $\hat{c}_{\bm x +\bm \mu_i^a,\sigma
}$, respectively. 
Here, $\bm x \in\mathscr{E}_d$ and $\bm \mu_i^a = \bm e_a/2+\tilde{\bm \mu}_i$.
Similarly to the main text, by applying a unitary transformation to the operators at sites in $\bigcup_{i = 1}^{N_{\rm sub}-1}\mathscr{I}_d^i$ with the common $a$ and $\bm x$, we introduce new operators satisfying the fermionic anticommutation relations: 
\begin{align}
    \hat{\bm \gamma}_{\bm x+\bm e_a/2} = \mathcal{U}_\theta \hat{\bm c}_{\bm x+\bm e_a/2,\sigma},
\end{align}
where $\hat{\bm \gamma}_{\bm x+\bm e_a/2,\sigma} = (\hat{\gamma}_{1,\bm x+\bm e_a/2,\sigma}\cdots\hat{\gamma}_{N_{\rm sub}-1,\bm x+\bm e_a/2,\sigma})^{\mathsf T}$ and $\hat{\bm c}_{\bm x+\bm e_a/2,\sigma}=(\hat{c}_{\bm x+\bm \mu_1^a,\sigma}\cdots\hat{c}_{\bm x+\bm\mu_{N_{\rm sub}-1}^a,\sigma})^{\mathsf T}$.
Here, $\mathcal{U}_\theta$ is the $(N_{\rm sub}-1)$ by $(N_{\rm sub}-1)$ unitary matrix and the continuum function of $\theta$.
We also define $\hat{\gamma}_{0,\bm x,\sigma}\equiv\hat{c}_{\bm x,\sigma}$.
Then, all of these $\gamma$ operators satisfy fermionic anticommutation relations.

By using these $\gamma$ operators, we construct a Hubbard Hamiltonian with tunable quantum geometry.
To this end, we introduce the following operators
\begin{align}
    \hat{\alpha}_{\bm x,\sigma} = \lambda\hat{\gamma}_{0,\bm x,\sigma}
    -\sum_{a}(\hat{\gamma}_{1,\bm x+\bm e_a/2,\sigma}+
    \hat{\gamma}_{1,\bm x-\bm e_a/2,\sigma}),\notag\\
    \hat{\beta}_{\bm x+\bm e_a/2,\sigma} = \hat{\gamma}_{0,\bm x,\sigma}+\lambda\hat{\gamma}_{1,\bm x+\bm e_a/2,\sigma}+\hat{\gamma}_{0,\bm x+\bm e_a,\sigma}.
\end{align}
Then, our noninteracting Hamiltonian with tunable quantum geometry is given by,
\begin{align}
    \hat{H}_0&=\hat{H}_{0:\rm T}+\hat{H}_{0:\rm ad}~\label{eq:H0_ex_Tasaki}\\
    \hat{H}_{0:\rm T}&=\sum_{\sigma}\sum_{\bm x}
    \left(
    -s\hat{\alpha}_{\bm x,\sigma}^\dagger\hat{\alpha}_{\bm x,\sigma}
    +t\sum_a\hat{\beta}_{\bm x+\bm e_a/2,\sigma}^\dagger\hat{\beta}_{\bm x+\bm e_a/2,\sigma}    
    \right),\\
    \hat{H}_{0:\rm ad} &= \sum_{a,i(\neq1),\sigma}\sum_{\bm x}v_{a,i}\hat{\gamma}_{i,\bm x + \bm e_a/2,\sigma}^\dagger \hat{\gamma}_{i,\bm x +\bm e_a/2,\sigma} + \hat{H}_{0:\rm arb}.
\end{align}
Here, $\hat{H}_{0:\rm ad}$ is arbitrary under the condition that $\hat{H}_{0:\rm arb}$ is a positive semidefinite Hermitian operator constructed only by the products of $\hat{\gamma}_{i(\neq 1),\bm x+\bm e_a/2,\sigma}$ and $\hat{\gamma}_{j(\neq 1),\bm x^\prime+\bm e_b/2,\sigma^\prime}^\dagger$.
For $d = 1$, by a suitable choice of $\hat{H}_{0:\rm arb}$ and $v_{a,i}$, $\hat{H}_0$ reduces to that defined in the main text.
By adding the on-site Coulomb repulsion, $\hat{H}_{\rm int} = U\sum_{\bm r\in\Lambda_d}\hat{n}_{\bm r,\uparrow}\hat{n}_{\bm r,\downarrow}$ with $\hat{n}_{\bm r,\sigma} = \hat{c}_{\bm r,\sigma}^\dagger\hat{c}_{\bm r,\sigma}$,
our Hubbard model is given by $\hat{\mathcal{H}} = \hat{H}_0+\hat{H}_{\rm int}$.
Since the $\gamma$ operators satisfy the fermionic anticommutation relations independently of $\theta$, the eigenvalues of $\hat{H}_{0}$ do not depend on $\theta$. However, the Bloch wavefunctions do depend on $\theta$ since the hopping processes vary with $\theta$. Thus, we can tune the quantum geometry of $\hat{H}$ independently of the energy dispersion and the Coulomb interaction, as discussed in the main text.

\subsection{Saturated ferromagnetism}

We sketch the proof of saturated ferromagnetism following Ref.~\cite{STasaki2020}. 
This proof is a straightforward extension of the one for the one-dimensional case presented
in the main text.
Hereafter, we assume $t,s,v_{a,i}>0$ and set the number of electrons to $N$ 
so that the lowest-band is half-filled.
First, we decompose the Hubbard Hamiltonian as
\begin{align}
    \hat{H} &= \xi\hat{H}_{\rm flat} + \hat{H}_{0:\rm arb} + \sum_{\bm x}\hat{h}_{\bm x},\\
    \hat{H}_{\rm flat} &=\sum_{a,\sigma}\sum_{\bm x}
    (t\hat{\beta}_{\bm x+\bm e_a/2,\sigma}^\dagger\hat{\beta}_{\bm x+\bm e_a/2,\sigma}   
    +\sum_{i(\neq0,1)}v_{a,i}\hat{\gamma}_{i,\bm x + \bm e_a/2,\sigma}^\dagger \hat{\gamma}_{i,\bm x +\bm e_a/2,\sigma})
    +\hat{H}_{\rm int}
    ,\\
    \hat{h}_{\bm x} &= 
    -s\sum_{\sigma}\hat{\alpha}_{\bm x,\sigma}^\dagger\hat{\alpha}_{\bm x,\sigma}
    +\dfrac{t^\prime}{2}\sum_{a,\sigma}
    \left[\hat{\beta}_{\bm x+\bm e_a/2,\sigma}^\dagger\hat{\beta}_{\bm x+\bm e_a/2,\sigma}
    +\hat{\beta}_{\bm x-\bm e_a/2,\sigma}^\dagger\hat{\beta}_{\bm x-\bm e_a/2,\sigma}
    \right]\notag\\
    &+\sum_{a,i(\neq0,1),\sigma}\dfrac{v_{a,i}^\prime}{2}\left[\hat{\gamma}_{i,\bm x + \bm e_a/2,\sigma}^\dagger \hat{\gamma}_{i,\bm x +\bm e_a/2,\sigma}
    +\hat{\gamma}_{i,\bm x - \bm e_a/2,\sigma}^\dagger \hat{\gamma}_{i,\bm x - \bm e_a/2,\sigma}
    \right]
    +(1-\kappa)U^\prime\hat{n}_{\bm x,\uparrow}\hat{n}_{\bm x,\downarrow}\notag\\
     &
     +\kappa\dfrac{U^\prime}{2d}\sum_{a}\left(\hat{n}_{\bm x+\bm e_a,\uparrow}\hat{n}_{\bm x+\bm e_a,\downarrow}
     +\hat{n}_{\bm x-\bm e_a,\uparrow}\hat{n}_{\bm x-\bm e_a,\downarrow}
     \right)
     +\sum_{a,i}\dfrac{U^\prime}{2}(\hat{n}_{\bm x+\bm \mu_i^{a},\uparrow}\hat{n}_{\bm x+\bm \mu_i^{a},\downarrow}
     +\hat{n}_{\bm x+\bm \mu_i^{a}-\bm e_{a},\uparrow}\hat{n}_{\bm x+\bm \mu_i^{a}-\bm e_{a},\downarrow}),
\end{align}
with $t^\prime = t-\xi$, $v_{a,i}^\prime = v_{a,i}-\xi$, $U^\prime = U-\xi$, and $0<\kappa<1$.
The noninteracting part of $\hat{H}_{\rm flat}$ has the lowest isolated flat band, whose linearly independent eigenstates are given by
$\hat{\alpha}_{\bm x,\sigma}^\dagger\ket{0}$.
We can prove the flat-band ferromagnetism of $\hat{H}_{\rm flat}$ by showing that $\hat{\alpha}_{\bm x,\sigma}^\dagger\ket{0}$ satisfies the connectivity condition of Mielke's theorem~\cite{SMielke1999}, which is a necessary and sufficient condition for the flat-band ferromagnetism in a half-filled lowest flat band.
Therefore, $\ket{\hat{\psi}_{\rm all\uparrow}} = \prod_{\bm x}\hat{\alpha}_{\bm x,\uparrow}^\dagger\ket{0}$ is the unique ground state of $\hat{H}_{\rm flat}$ up to the trivial degeneracy due to the $SU(2)$ symmetry.
Then, since $\ket{\hat{\psi}_{\rm all\uparrow}}$ is one of the ground states of $\hat{H}_{0:\rm arb}$, i.e. $\hat{H}_{0:\rm arb}\ket{\hat{\psi}_{\rm all\uparrow}}=0$, $\ket{\hat{\psi}_{\rm all\uparrow}}$ is also the unique ground state of $\xi\hat{H}_{\rm flat} + \hat{H}_{0:\rm arb}$.

Because of $\hat{h}_{\bm x}\ket{\hat{\psi}_{\rm all\uparrow}} = -s(\lambda^2+2d)\ket{\hat{\psi}_{\rm all\uparrow}}$,  if the minimum eigenvalue of $\hat{h}_{\bm x}$ is $-s(\lambda^2+2d)$, 
then $\ket{\hat{\psi}_{\rm all\uparrow}}$ is the unique ground state of the total Hamiltonian, $\hat{H}$.
In other words, when $\bra{\hat{\Phi}}\hat{h}_{\bm x}\ket{\hat{\Phi}}\geq -s(\lambda^2+2d)$ for any normalized state $\ket{\hat{\Phi}}$, $\hat{H}$ exhibits saturated ferromagnetism.
For $t,v_{a,i}, U\rightarrow\infty$, we can analytically prove $\bra{\hat{\Phi}}\hat{h}_{\bm x}\ket{\hat{\Phi}}\geq -s(\lambda^2+2d)$.
Here, we show the basic strategy of the proof, which is based on Ref.~\cite{STasaki1995} and the main text.
When the total spin of $\ket{\hat{\Phi}}$ is maximum~\cite{totalspin}, the problem is reduced to a one-particle problem, and we can show $\bra{\hat{\Phi}}\hat{h}_{\bm x}\ket{\hat{\Phi}}\geq -s(\lambda^2+2d)$ by calculating the single-particle spectrum of $\hat{h}_{\bm x}$.
Therefore, we only focus on states whose total spin does not take the maximum value.
Clearly, it suffices to consider states such that $\lim_{t,v_{z,i},U\uparrow\infty}\bra{\hat{\Phi}}\hat{h}_{\bm x}\ket{\hat{\Phi}}<\infty$, which is referred to as the finite-energy condition.
We can show that the states, which are made up of the product of $\lambda\hat{\gamma}_{0,\bm x\pm\bm e_a,\sigma}^\dagger - \hat{\gamma}_{1,\bm x\pm\bm e_a/2,\sigma}^\dagger$ and do not contain a doubly occupied site, satisfy the finite-energy condition.
Then, following the proof in Sec. 11.4 of Ref.~\cite{STasaki2020}, one finds that such a state satisfies $\bra{\hat{\Phi}}\hat{h}_{\bm x}\ket{\hat{\Phi}}\geq -s(\lambda^2+2d)$ when $0<1/\lambda<f(\lambda)$ with $f(\lambda) = \sqrt{(2d+\sqrt{4d^2+12d-7})/4(d-1)}$.
Thus, when $0<1/\lambda<f(\lambda)$, the ground state of $\hat{H}$ with $t,v_{a,i}, U\rightarrow\infty$ exhibits saturated ferromagnetism.

\section{Stoner magnetism for multi-band systems}
In this section, we introduce Stoner's mean-field theory~\cite{SStoner1938-he} for multi-band systems and analyze the magnetism based on it.

\subsection{Mean-field theory}
First, we introduce the mean-field theory for a multi-band Hubbard Hamiltonian.
\subsubsection{Model and mean-field Hamiltonian}
We consider the multi-band Hubbard Hamiltonian: 
\begin{align}
  \hat{H} &= \hat{H}_0 + \hat{H}_{\rm int},\\
  \hat{H}_0 &= \sum_{ij}\sum_{\bm k}\hat{c}_{i}^\dagger(\bm k)h_{ij}(\bm k)\hat{c}_{j}(\bm k),\\
  \hat{H}_{\rm int} &= \sum_{ijkl}\sum_{\bm k,\bm k^\prime, \bm q}\Gamma_{ij;lk}
  \hat{c}_{i}^\dagger(\bm k)\hat{c}_{j}(\bm k + \bm q)
  \hat{c}_{k}^\dagger(\bm k^\prime)\hat{c}_{l}(\bm k^\prime-\bm q).
\end{align}
Here, $\hat{c}_{i}(\bm k)$ is the fermionic annihilation operator with the internal degrees of freedom $i$ and the momentum $\bm k$, where we use $i=(\rho_i,\sigma_i)$ as a composite index for the sublattice (and orbital) $\rho_i$ and the spin $\sigma_i$.
$\Gamma_{ij;lk}^* = \Gamma_{lk;ij}$ is satisfied since $\hat{H}_{\rm int}$ is Hermitian.
By applying a mean-field decomposition to $\hat{H}_{\rm int}$ and ignoring constant terms, we obtain
\begin{align}
   \hat{H}_{\rm int} 
  &\approx\sum_{ij}\sum_{\bm k,\bm q}(M_{ij}(\bm q)+M_{ji}^*(-\bm q))
  \hat{c}_{i}^\dagger(\bm k)\hat{c}_{j}(\bm k-\bm q),
\end{align}
with
\begin{align}
  M_{ij}(\bm q) &\equiv \sum_{kl}\sum_{\bm k}\Gamma_{ij;lk}
  \braket{\hat{c}_{k}^\dagger(\bm k)\hat{c}_{l}(\bm k+\bm q)},~\label{eq:mean1}\\
  M_{ji}^*(-\bm q) &\equiv \sum_{kl}\sum_{\bm k}\Gamma_{kl;ji}
  \braket{\hat{c}_{k}^\dagger(\bm k)\hat{c}_{l}(\bm k+\bm q)},~\label{eq:mean2}
\end{align}
where $\braket{\cdots}$ denotes the thermal average with respect to the mean-field Hamiltonian, which will be defined below. 
By introducing the molecular field
\begin{align}
    \bar{h}_{ij}(\bm q) \equiv M_{ij}(\bm q)+M_{ji}^*(-\bm q),
\end{align}
the mean-field Hamiltonian is given by
\begin{align}
    \hat{H}_{\rm MF} &= \sum_{ij}\sum_{\bm k_1,\bm k_2}\hat{c}_{i}^\dagger(\bm k_1)h_{ij}^{\rm MF}(\bm k_1,\bm k_2)\hat{c}_{j}(\bm k_2),~\label{eq:Hmf}\\
  h_{ij}^{\rm MF}(\bm k_1,\bm k_2) &= \delta_{\bm k_1,\bm k_2}h_{ij}(\bm k_1) + \bar{h}_{ij}(\bm k_1-\bm k_2).
\end{align}

\subsubsection{Linearized mean-field equation}
Here, we derive the linearized mean-field equation for Eqs.~\eqref{eq:mean1} and ~\eqref{eq:mean2}.
We introduce the Green function for Eq.~\eqref{eq:Hmf} as
\begin{align}
    G_{ij}^{\rm MF}(\bm k_1,\bm k_2;i\omega_n) &= -\int_0^\beta d\tau e^{i\omega_n\tau}
    \braket{T_{\tau}[\hat{c}_{i}(\bm k_1,\tau)\hat{c}_{i}^\dagger(\bm k_2)]}
\end{align}
where $T_{\tau}$ is the time-ordering operator for imaginary time $\tau$ and $\omega_n$ is the fermionic Matsubara frequency. 
We adopt natural units with $k_{\mathrm B}=1$, so that the inverse temperature is given by $\beta = 1/T$.
We write the matrix representation of $G^{\rm MF}_{ij}(\bm k_1,\bm k_2;i\omega_n)$ as $G^{\rm MF}(\bm k_1,\bm k_2;i\omega_n)$.
Practically, the Green function is calculated via an explicit formula for the matrix elements of the inverse matrix 
\begin{align}
    [G^{\rm MF}(\bm k_1,\bm k_2;i\omega_n)]^{-1}_{ij} &= (i\omega_n + \mu_{\rm c})\delta_{\bm k_1,\bm k_2}\delta_{ij} - h_{ij}^{\rm MF}(\bm k_1,\bm k_2),
\end{align}
where $\mu_{\rm c}$ is the chemical potential.
We rewrite the mean-field expectation value in terms of the Green function as
\begin{align}
    \braket{\hat{c}_{k}^\dagger(\bm k)\hat{c}_{l}(\bm k+\bm q)} 
  &=\dfrac{1}{\beta}\sum_{\omega_n} e^{i\omega_n0^+} G_{lk}^{\rm MF}(\bm k+\bm q,\bm k;i\omega_n),
\end{align}
where $0^+$ indicates an infinitesimal positive number.
By using the Green function, Eqs.~\eqref{eq:mean1} and~\eqref{eq:mean2} are given by
\begin{align}
  M_{ij}(\bm q) &= \dfrac{1}{\beta}\sum_{kl}\sum_{\bm k,i\omega_n} e^{i\omega_n0} \Gamma_{ij;lk}G_{lk}^{\rm MF}(\bm k+\bm q,\bm k;i\omega_n),~\label{eq:mean_green1}\\
  M_{ji}^*(-\bm q) &= \dfrac{1}{\beta}\sum_{kl}\sum_{\bm k,i\omega_n} e^{i\omega_n0} \Gamma_{kl;ji}G_{lk}^{\rm MF}(\bm k+\bm q,\bm k;i\omega_n).~\label{eq:mean_green2}    
\end{align}
These expressions explicitly show their equivalence.
We expand the Green function to first order in the molecular field as
\begin{align}
  G_{lk}^{\rm MF}(\bm k_1,\bm k_2;i\omega_n) &= \delta_{\bm k_1,\bm k_2}G_{lk}(\bm k;i\omega_n)
  +\sum_{mn}G_{0:lm}(\bm k_1;i\omega_n)\bar{h}_{mn}(\bm k_1-\bm k_2)G_{nk}(\bm k_2;i\omega_n)
  +O(\bar{h}^2),~\label{eq:Gexpand}
\end{align}
Here, the noninteracting Green function, $G_{0:ij}(\bm k;i\omega_n)$, is defined via its matrix representation, $G_0(\bm k;i\omega_n)$, as
\begin{align}
    [G_0(\bm k;i\omega_n)]^{-1}_{ij}&=(i\omega_n-\mu_{\rm c})\delta_{ij} - h_{ij}(\bm k).
\end{align}
By inserting Eq.~\eqref{eq:Gexpand} into Eqs.~\eqref{eq:mean_green1} and~\eqref{eq:mean_green2}, to first order in $\bar{h}$, we obtain
\begin{align}
    M_{ij}(\bm q) &= \Gamma_{ij;lk}
  (\delta_{\bm q,0}n_{lk}-\chi_{lk,mn}(\bm q)\bar{h}_{mn}(\bm q)),~\label{eq:mean1_linear}\\
  M_{ji}^*(-\bm q) &= \Gamma_{ij;lk}^*
  (\delta_{\bm q,0}n_{lk}-\chi_{lk,mn}(\bm q)\bar{h}_{mn}(\bm q)),~\label{eq:mean2_linear}\\
  n_{lk} &\equiv \dfrac{1}{\beta}\sum_{\bm k,\omega_n} e^{i\omega_n0} G_{lk}(\bm k;i\omega_n),\\
  \chi_{lk,mn}(\bm q) &\equiv -\dfrac{1}{\beta}\sum_{\bm k,\omega_n} e^{i\omega_n0}
  G_{lm}(\bm k+\bm q;i\omega_n)G_{nk}(\bm k;i\omega_n).
\end{align}
In the following, we implicitly sum over repeated indices. 
By summing up Eqs.~\eqref{eq:mean1_linear} and~\eqref{eq:mean2_linear}, we obtain a linearized mean-field equation,
\begin{align}
  \bar{h}_{ij}(\bm q) 
  &= (\Gamma_{ij;kl}+\Gamma_{ij;kl}^*)
  (\delta_{\bm q,0}n_{kl}-\chi_{lk,mn}(\bm q)\bar{h}_{mn}(\bm q)).
\end{align}
This leads to
\begin{align}
    [\bm 1+ (\Gamma+\Gamma^*)\chi(\bm q)]\bar{h}(\bm q)
    =\delta_{\bm q,0}(\Gamma+\Gamma^*)n,
\end{align}
with identity matrix $\bm 1$ where $\Gamma$ and $\chi(\bm q)$ are matrix representations of $\Gamma_{ij;kl}$ and $\chi_{ij;kl}(\bm q)$, respectively, and $\bar{h}(\bm q)$ and $n$ are the vector representations of $\bar{h}_{ij}(\bm q)$ and $n_{ij}$, respectively.

\subsection{Stoner criterion for spin-independent hoppings and onsite Coulomb interaction}
We consider the $SU(2)$-symmetric Hamiltonian with spin-independent hopping and Coulomb interaction, i.e.,
\begin{align}
     h_{ij}(\bm k)&= \delta_{\sigma_i,\sigma_j}h_{\rho_i\rho_j}^{(0)}(\bm k)\label{eq:h0ij},\\
    \Gamma_{ij;kl} &= \dfrac{U}{2}\delta_{i,j}\delta_{k,l}
    \delta_{\rho_i,\rho_k}(1-\delta_{\sigma_i,\sigma_k}).
\end{align}
Then, the linearized mean-field equation is rewritten as
\begin{align}
  \bar{h}_{i}(\bm q) &= (1-\delta_{\sigma_i,\sigma_j})\delta_{\rho_i,\rho_j}U
  \left[\delta_{\bm q,0}n_{j}-\chi_{j,k}(\bm q)\bar{h}_{k}(\bm q)\right],~\label{eq:linear_mean_su2}
\end{align}
with abbreviated notation $\bar{h}_{i}(\bm q) \equiv\bar{h}_{ii}(\bm q)$, $n_{0:i}\equiv n_{0:ii}$, and $\chi_{0:i,j}(\bm q)\equiv\chi_{0:ii,jj}(\bm q)$.
The magnetic molecular field is defined by
\begin{align}
    \bar{h}_{z;\rho_i}(\bm q) &= \bar{h}_{(\rho_i,\uparrow)}(\bm q)-\bar{h}_{(\rho_i,\downarrow)}(\bm q).
\end{align}
By using Eq.~\eqref{eq:linear_mean_su2}, this is rewritten by,
\begin{align}
    \bar{h}_{z;\rho_i}(\bm q)&=-U\chi_{0:\rho_i,\rho_j}(\bm q)(\bar{h}_{(\rho_j,\downarrow)}(\bm q)-\bar{h}_{(\rho_j,\uparrow)}(\bm q))\notag\\
    &=U\chi_{0:\rho_i,\rho_j}(\bm q)\bar{h}_{z;\rho_j}(\bm q),
\end{align}
with $\chi_{0:\rho_i,\rho_j}(\bm q) \equiv \chi_{(\rho_i,\uparrow),(\rho_j,\uparrow)} (\bm q)=\chi_{(\rho_i,\downarrow),(\rho_j,\downarrow)}(\bm q)$.
As a result, we obtain
\begin{align}
    [\bm 1-U\chi_{0}(\bm q)]\bar{h}(\bm q) = 0.
\end{align}
This indicates that magnetic order with $\bar{h}(\bm q)\propto \xi(\bm q)$ emerges when the largest eigenvalue of $U\chi_{0}(\bm q)$ reaches unity, where $\xi(\bm q)$ is the corresponding eigenvector.
Thus, the Stoner criterion for multi-band systems is given 
\begin{align}
    U\bar{\chi}_0(\bm q)\geq 1,
\end{align}
where $\bar{\chi}_0(\bm q)$ is the maximum eigenvalue of $\chi(\bm q)$.
When this criterion is satisfied, the corresponding magnetic order is stabilized.

\subsection{Application to 1D extended Tasaki lattice}
We study the magnetism of the Hubbard model on a 1D extended Tasaki lattice based on Stoner theory.
We adopt the Hubbard Hamiltonian defined in Sec.~\ref{sec:Hubbar_qg} for $d = 1, N_{\rm sub} = 5$ and 
\begin{align}
    \mathcal{U}_\theta = \left(
    \begin{array}{cccc}
       \cos^2\theta &
       \frac{1}{2}\sin2\theta  &
       \frac{1}{2}\sin2\theta & \sin^2\theta \\  
       -\sin\theta & \cos\theta & 0 & 0 \\  
       -\frac{1}{2}\sin2\theta & -\sin^2\theta & \cos^2\theta & \frac{1}{2}\sin2\theta \\  
       0 & 0 & -\sin\theta & \cos\theta
    \end{array}
    \right).
\end{align}
We choose $v_{a,i}$ and $\hat{H}_{\rm 0:arb}$ such that the resulting Hamiltonian reproduces 
the Hubbard Hamiltonian analyzed in the main text; all parameters are set to the same values as those used in the main text.
Also, the chemical potential is adjusted so that the lowest band is half-filled.

The fermionic operators in momentum space are defined as
\begin{align}
    \hat{c}_{i,\sigma}(k) = \frac{1}{\sqrt{N}}\sum_{k}e^{-ikr}\hat{c}_{r+\mu_i,\sigma}~\label{eq:fourier_1DTasaki}
\end{align}
for $i = \{0,\cdots,N_{\rm sub} - 1\}$ with $\mu_0 = 0$.
Since we consider the 1D system, we omit the $a$ dependence of $\mu_i^a$, and $k$, $r$ and $\mu_i$ are scalar quantities and are denoted by non-bold characters.
By combining Eq.~\eqref{eq:fourier_1DTasaki} with Eq.~\eqref{eq:H0_ex_Tasaki}, we obtain the noninteracting Hamiltonian in momentum space where the coefficient of $\hat{c}_{i,\sigma}^\dagger(k)\hat{c}_{j,\sigma}(k)$ corresponds to $h_{i,j}^{(0)}(k)$.

\begin{figure}[tbp]
  \includegraphics[width=1.0\linewidth]{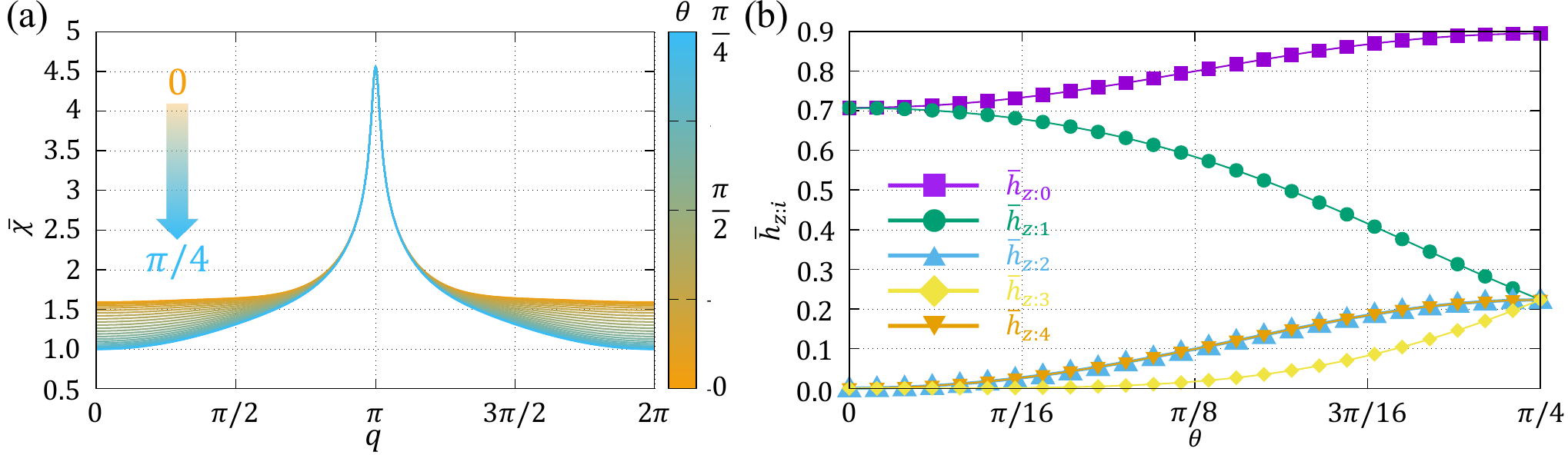}
  \centering
  \caption{(a)The $\bm q$ dependence of  $\bar{\chi}(\bm q)$ for several values of $\theta$. The blue and orange lines correspond to $\theta = 0$ and $\pi/4$, respectively.
  (b) $\theta$ dependence of $\bar{h}_{z:i}(0)$. The purple square, the green circle, the blue triangle, the yellow diamond, and the orange inverse triangle correspond to $h_{z:0}, h_{z:1}, h_{z:2}, h_{z:3}$, and $h_{z:4}$, respectively.
  }
  \label{fig:stoner}
\end{figure}

The numerical evaluation of $\bar{\chi}_0(\bm q)$ at zero temperature is computationally demanding.
Therefore, we perform calculations at $T=0.001$. 
In Fig.~\ref{fig:stoner}, we show the $\bm q$ dependence of $\bar{\chi}(\bm q)$ for several values of $0\leq\theta\leq\pi/4$.
It is found that the $\bar{\chi}(0)$ corresponding to ferromagnetic order is not the maximum; instead, $\bar{\chi}(\pi)$ attains the maximum, indicating that antiferromagnetism is stable within the Stoner theory. Thus, we conclude that Stoner's mean-field theory fails to reproduce our results in the main text based on the rigorous theory and the spin-wave theory.
At $q = \pi$, the molecular field is given by $\bar{h}_{z:i}(\pi) = \delta_{i,0}$.
In other words, the antiferromagnetism occurs at the sites in $\mathscr{E}_1$ within the Stoner theory.

On the other hand, the $\theta$ dependence of $\bar{\chi}(0)$ is consistent with the results in the main text, where we demonstrated that saturated ferromagnetism becomes unstable as $\theta$ increases; correspondingly, $\bar{\chi}(0)$ decreases with increasing $\theta$.
Moreover, the $\theta$ dependence of the molecular field is consistent with the intuitive physical picture of saturated ferromagnetism.
Intuitively, saturated ferromagnetism is stabilized by the overlap of Wannier states, which occur at the sites in $\mathscr{I}_1^i$ and generate an effective ferromagnetic exchange coupling.
Thus, as $\theta$ increases, it is expected that the overlaps between Wannier states of the lowest band decrease, and the resulting effective ferromagnetic exchange coupling is reduced, leading to the instability of the ferromagnetic state.

Within Stoner theory, this mechanism is reflected in the molecular field.
In Fig.~\ref{fig:stoner}(b), we show the $\theta$ dependence of $\bar{h}_{z:i}(0)$.
At $\theta = 0$, the Wannier states of the lowest band are localized only on the sites in $\mathscr{E}_1$ and $\mathscr{I}_1^1$, which are decoupled from the other sites; correspondingly, only the molecular fields on these sites are finite. As $\theta$ increases, these sites become hybridized with the other sites, and sites other than $\mathscr{E}_1$ and $\mathscr{I}_1^1$ also contribute to the Wannier states.
Then, $\bar{h}_{z:i}$ for the sites other than $\mathscr{E}_1$ and $\mathscr{I}_1^1$ become finite.
However, because of this mixing, the maximum value of $\bar{h}_{z:i}$ on the sites that support the overlaps (sites in $\mathscr{I}_1^i$) is suppressed.
In other words, the reduction of the overlaps between Wannier states is directly encoded in the suppression of the maximum molecular field of the sites supporting the overlaps between Wannier states.

\begin{figure}[tbp]
  \includegraphics[width=1.0\linewidth]{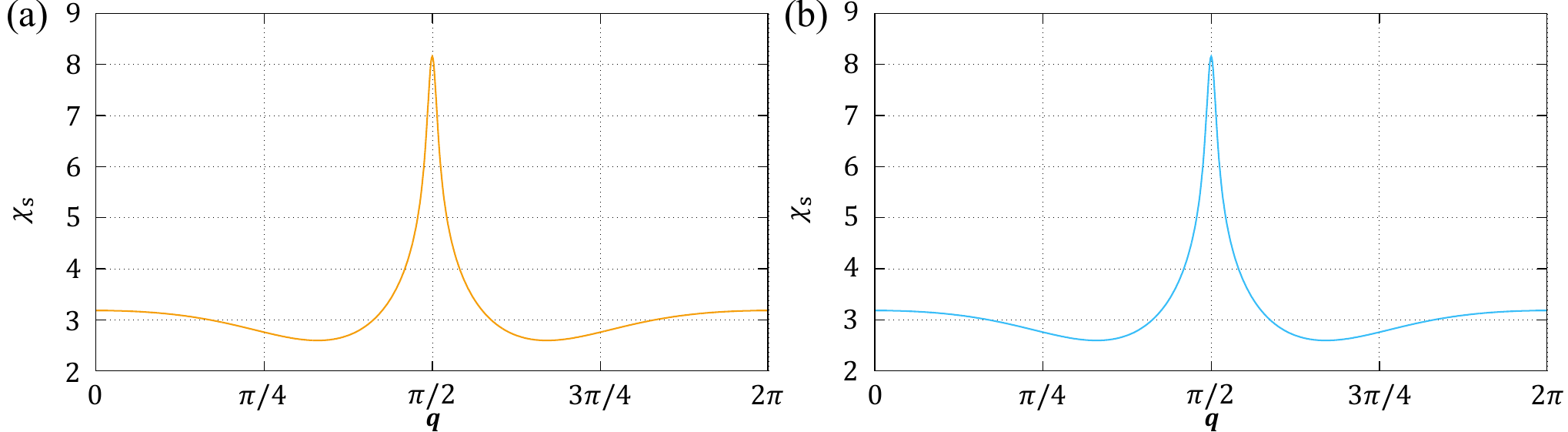}
  \centering
  \caption{$\chi_{\rm s}(\bm q)$
  for (a) $\theta= 0$ and (b) $\theta = \pi/4$.
  }
  \label{fig:chis}
\end{figure}

We also emphasize that the $\theta$-dependence of $\bar{\chi}(\bm q)$ is not given by the canonical spin susceptibility for magnetic dipole, i.e., $\chi_{\rm s}(\bm q) = \sum_{ij}\chi_{0:ij}(\bm q)$.
By using the numerical calculation, we confirm that $\chi_{\rm s}(\bm q)$ is independent of $\theta$.
For reference, we show $\chi_{\rm s}(\bm q)$ of $\theta= 0$ and $\theta= \pi/4$ in Figs.~\ref{fig:chis} (a) and (b), respectively.
Clearly, the two profiles are identical. 
This result indicates that higher-order magnetic multipoles other than the magnetic dipole are important in the quantum-geometry-driven magnetic phase transition in our model.


\begin{thebibliography}{108}
\bibitem [{deg()}]{degenerate}%
  \BibitemOpen
  \href@noop {} {}\bibinfo {note} {For simplicity, we set the hopping
  parameters in the different chains to the same value, but this is not crucial
  for the subsequent results. These degeneracies can easily be lifted by using
  different hopping values for the different chains.}\BibitemShut {Stop}%
\bibitem [{uni()}]{unitary}%
  \BibitemOpen
  \href@noop {} {}\bibinfo {note} {Restricting $\mathcal{U}_{\theta}$ to
  real matrices continuously connected to the identity gives
  $\mathcal{U}_{\theta}\in\mathrm{SO}(4)$. The six elementary rotations
  $e^{\theta_{ij}J_{ij}}$ generate $\mathrm{SO}(4)$, where
  $[J_{ij}]_{a,b}=\delta_{a,i}\delta_{b,j} - \delta_{a,j}\delta_{b,i}$ defines
  the generators of the Lie algebra $\mathfrak{so}(4)$. Since each factor
  $e^{\theta_{ij}J_{ij}}$ connects only the $i$th and $j$th chains, at least
  three factors are needed to connect all four chains. Eq.~\eqref{eq:Utheta}
  corresponds to a minimal choice $\mathcal{U}_{\theta}=e^{\theta
  J_{13}}e^{\theta J_{12}}e^{\theta J_{34}}$ with a common angle
  $\theta$.}\BibitemShut {Stop}%
\bibitem [{suf()}]{sufficient}%
  \BibitemOpen
  \href@noop {} {}\bibinfo {note} {Let $E_0$ be the ground state energy of
  $\hat{H}$. While $N\varepsilon_0$ gives a lower bound on $E_0$, i.e., $E_0
  \geq N\varepsilon_0$, the equality does not necessarily hold.}\BibitemShut
  {Stop}%
\end{thebibliography}
\begin{thebibliography}{15}
\bibitem [{var()}]{variable}%
  \BibitemOpen
  \href@noop {} {}\bibinfo {note} {One way to see this rearrangement of $\bm
  k,\bm k^\prime$, and $\bm q$ is to transform the interaction back to real
  space. Then, performing the Fourier transformation again and taking the
  appropriate variables yields the third-line expression.}\BibitemShut {Stop}%
\bibitem [{tot()}]{totalspin}%
  \BibitemOpen
  \href@noop {} {}\bibinfo {note} {Total spin operator is defined by $\hat{\bm
  S}_{\rm tot}= \sum_{\bm r}\hat{\bm c}_{\bm r}^\dagger \bm \sigma \hat{\bm
  c}_{\bm r}/2$ with $\hat{\bm c}_{\bm r} = (\hat{c}_{\bm r,\uparrow},
  \hat{c}_{\bm r,\downarrow})$. Because of the $SU(2)$ symmetry, the eigenstate
  of $\hat{h}_{\bm x}$ is also the eigenstate of $(\hat{S}_{\rm tot})^2$ with
  the eignenvalue $S_{\rm tot}(S_{\rm tot}+1)$. The total spin is characterized
  by $S_{\rm tot}$.}\BibitemShut {Stop}%
\end{thebibliography}
\end{document}